\documentclass[11pt,a4paper]{article}

\usepackage[T1]{fontenc}
\usepackage[utf8]{inputenc}
\usepackage{lmodern}
\usepackage[brazilian,english]{babel}
\usepackage[margin=2.6cm]{geometry}
\usepackage{microtype}
\usepackage{array}
\usepackage{booktabs}
\usepackage{tabularx}
\usepackage{ragged2e}
\usepackage[font=small,labelfont=bf]{caption}
\usepackage[hidelinks]{hyperref}
\usepackage{xurl}
\usepackage{enumitem}

\setlist{itemsep=2pt,topsep=4pt}
\newcolumntype{L}{>{\RaggedRight\arraybackslash}X}

\newenvironment{abntrefs}{%
  \par\begingroup\small\setlength{\parindent}{0pt}%
  \setlength{\parskip}{0.55em}\RaggedRight}{\par\endgroup}

\newcommand{\refavail}{Available at:}
\newcommand{\refaccess}{Accessed:}
\newcommand{\refoct}{Oct.}
\newcommand{\refdec}{Dec.}
\newcommand{\refnov}{Nov.}
\newcommand{\refreissue}{Reissue:}
\newcommand{\refcatalog}{Catalogue record:}

\newenvironment{aidisclaimer}[1]{%
  \par\medskip\noindent\begin{lrbox}{0}\begin{minipage}{\dimexpr\linewidth-2\fboxsep-2\fboxrule\relax}%
  \small\textbf{#1.}\ }{\end{minipage}\end{lrbox}\fbox{\usebox{0}}\par\medskip}

\newcommand{\reading}[1]{\textit{#1}}

\hypersetup{
  pdftitle={The user who is not there: a genealogy and critique of simulated users in UX testing},
  pdfauthor={Jose Carlos Porto Arcoverde Jr.},
  pdfkeywords={usability testing, synthetic users, large language models, participatory design, labor process, research ethics, Latin America}
}

\title{The user who is not there:\\[0.3em]
\large a genealogy and critique of simulated users built with generative AI in UX testing}

\author{José Carlos Porto Arcoverde Jr. (Mabuse)\\[0.3em]
\small Graduate Program in Design, Universidade Federal de Pernambuco (PPGDesign-UFPE)\\
\small CESAR School, Recife, Brazil\\
\small \texttt{josecarlos.arcoverde@ufpe.br}}

\date{November 2026}

\begin{document}

\maketitle

\begin{center}
\small\textit{Language note.} The original language of this paper is Portuguese.
Part I is the English version. Part II is the Portuguese original (p.~\pageref{pt-start}).
\end{center}

\begin{abstract}
\noindent
Usability testing depends on people, and people's time does not speed up at the pace that software production demands. Large language models (LLMs) now promise to remove people from testing by generating simulated users that browse websites, answer questionnaires and give interviews. This paper examines how academic research understands that promise. It combines a critical literature review with a genealogical reading that follows the simulated user from the cognitive models of the 1980s to current generative agents. The paper makes four claims. First, testing without the user is an old aim of human-computer interaction (HCI), and cost has always justified it. Second, the validity criterion for user models shifted from predictive precision to believability, and recent studies that return to precision report weak results. Third, simulated participants conflict with the values of academic research and with methods that treat participation as a political right, with Participatory Design as the limit case. Fourth, the Latin American literature found in this review uses generative AI to support expert inspection and contains no study of synthetic users. That gap leaves open questions about who gets represented and about research ethics regulation in Brazil. The paper closes with an agenda of open problems. As part of its method, it reports that the review itself used an LLM as research assistant, and it discusses what that implies.

\medskip
\noindent\textbf{Keywords:} usability testing; synthetic users; large language models; participatory design; labor process; research ethics; Latin America.
\end{abstract}

\begin{aidisclaimer}{Declaration on the use of generative AI}
This paper was prepared with the help of a large language model (Claude, by Anthropic, in versions available between October and November 2026), used as a research assistant. The model searched the literature, summarised sources, checked bibliographic records, drafted the Portuguese text and translated it into English. The research questions, the theoretical framing and the editorial decisions are the author's. Every reference was checked against at least one primary record, and Section~\ref{sec:method} states the limits of that check. The model is not an author of this paper. The author takes full responsibility for its content, including any errors. Sections~\ref{sec:ai-use} and~\ref{sec:reflexive} describe and discuss this process.
\end{aidisclaimer}

\part*{Part I: English version}
\addcontentsline{toc}{part}{Part I: English version}

\section{Introduction}\label{sec:intro}

Usability testing consults people. It places someone in front of an interface, asks for a task and observes what happens. That consultation takes time. Teams must recruit participants, schedule sessions, run them and analyse the results. For this reason, human-computer interaction (HCI) has long looked for ways to evaluate interfaces without users.

Generative AI offers the most radical version of that search so far. A large language model (LLM) can generate thousands of simulated users. Each one has a persona, a memory and a way of speaking. These agents browse websites, answer questionnaires and give interviews. The literature calls them synthetic users, silicon samples or simulated agents (ARGYLE \textit{et al.}, 2023; LU \textit{et al.}, 2025).

The promise raises three questions. The first is epistemological: what can a simulated user tell us about real people? The second is ethical: what is lost, and who is harmed, when the person leaves the research? The third is political: whom does the acceleration offered by simulation serve?

This paper addresses those questions with a critical literature review organised as a genealogy. Its contributions are the following:
\begin{itemize}
  \item a genealogy of the simulated user in HCI, from cognitive models to generative agents, which identifies a shift in the validity criterion for user models (Section~\ref{sec:genealogy});
  \item a reading of synthetic users as a response to the time pressure of capitalist software production, based on labour process theory (Section~\ref{sec:problem});
  \item an analysis of the conflicts that synthetic participants create with academic research and with Participatory Design (Sections~\ref{sec:problem} and~\ref{sec:approaches});
  \item a survey of Latin American work on the topic, which found no study of synthetic users (Section~\ref{sec:latam});
  \item an agenda of open problems (Section~\ref{sec:future}).
\end{itemize}

Section~\ref{sec:method} describes the method, including the use of an LLM as research assistant. Section~\ref{sec:discussion} discusses the findings and Section~\ref{sec:conclusions} concludes.

Throughout the paper, passages that present this paper's own interpretation, rather than a claim made by the cited sources, are marked as such in the text.

\section{Method}\label{sec:method}

\subsection{Approach}

The paper combines two procedures. The first is a critical review of research on LLM-based simulation of users and evaluators in usability and interaction studies. The second is a genealogical reading of the idea of a simulated user. That reading asks what prepared the idea, either by affinity or by conflict. It then asks how the idea was formulated and how later work contested or revisited it. A genealogy does not assume a straight line of progress. It looks for disputes, breaks and returns.

\subsection{Search and selection}

Searches ran in October 2026. The policy information in Section~\ref{sec:ai-use} was updated in November 2026. Sources included the ACM Digital Library, arXiv, the ACL Anthology, publisher pages (Cambridge University Press, SAGE, Springer, MIT Press), the SBC OpenLib repository of the Brazilian Computer Society, university repositories and the dblp database. Queries ran in English, Portuguese and Spanish. They combined terms on usability, user experience, heuristic evaluation, synthetic users, LLM agents, research participants and personas with terms on Participatory Design, labour and coloniality. Reference lists of key papers were followed to find further work.

A work was included when it addressed at least one of these topics: simulation of users or evaluators with LLMs; critiques of replacing human participants with AI; historical antecedents of user modelling and automated evaluation; or the theoretical frames used in the analysis. The final corpus has 42 works.

\subsection{Use of generative AI in this study}\label{sec:ai-use}

This paper was produced through an extended dialogue between the author and an LLM (Claude, by Anthropic) used as a research assistant. The model ran web searches, summarised publisher pages and abstracts, checked bibliographic records, drafted text in Portuguese and produced the English version. The author set the research questions and the theoretical framing. The framing through the mode of production, the treatment of Participatory Design and the search for Latin American work all came from the author's requests during the dialogue. The author takes full responsibility for the content.

This disclosure is part of the method and not only a formality. A paper that criticises the replacement of research labour by LLMs was itself produced with an LLM. Section~\ref{sec:reflexive} discusses what that means.

\subsection{Verification and limitations}

Every reference was checked against at least one primary record: a publisher page, a repository or a bibliographic database. A second check, in November 2026, re-examined all 42 references. Each one now has a DOI, a publisher or repository link, or an ISBN with a catalogue record. One claim that could not be traced to its source was removed. Books are cited for their central thesis, without page numbers. One reference is a preprint not yet peer reviewed (ROTHSCHILD \textit{et al.}, 2025), and the text marks it as such.

The review has limits that the reader should keep in mind. Many sources were assessed through abstracts, publisher summaries and partial text, not through full reading. Claims about those sources stay within what such material supports. Searches in Spanish returned mostly non-academic material. The review did not cover traditions such as activity ergonomics, which could offer other critiques of the simulated user.

\section{The problem: production, knowledge and participation}\label{sec:problem}

\subsection{A need of the mode of production}

In software production, usability testing is a stage in the cycle that takes a product from conception to market. It is a slow stage, because it depends on the time of other people.

Under capitalist competition, the time between conceiving and selling is a cost. The firm that launches first reaches the market first. Any stage that cannot be accelerated becomes an obstacle to the valorisation of capital. Research with users is one of those stages. It does not speed up at the pace of code, because the person consulted does not work at the pace of the machine. \reading{This is the paper's reading.} The technical literature names the same fact in other words. Ivory and Hearst (2001, p. 470) justified automated usability evaluation because evaluation ``can be expensive in terms of time and human resources''. Agnew \textit{et al.} (2024) found that reducing research costs is one of the main motivations behind proposals to replace human participants with generative AI.

Synthetic users answer that pressure directly. They work at machine pace, at any hour and in any number.

\subsection{Separating conception from execution}

Braverman (1974) showed that monopoly capitalism reorganises the labour process in two moves. The first separates the conception of work from its execution. The second removes knowledge of the craft from the worker and transfers it to management and to machines. Qin and Cheon (2026) proposed bringing labour process theory into HCI. They examine the HCI and computer-supported cooperative work literature on labour, and they ask the field to treat labour as a question of political economy, not only as task performance with technology.

\reading{This paper reads the simulated user as a repetition of Braverman's two moves, on two planes.} On the first plane, it removes knowledge from the participant. What a person knows about how they use, avoid and work around an interface turns into a statistical pattern extracted from texts that the person wrote for other purposes. On the second plane, it removes knowledge from the UX researcher. Running a session, listening, and reading silence and hesitation turn into an automated procedure. What once required a craft now requires a service subscription.

The movement also reproduces itself. User research had already passed through a first round of commodification with crowd work platforms such as Amazon's Mechanical Turk. Gray and Suri (2019) describe the invisible human labour behind such automated services. Kapania \textit{et al.} (2025) note that LLMs were trained on texts written by people who did not consent to that use and on the work of poorly paid data annotators. The simulation of those people is then offered as a product. Westwood (2025) showed that an autonomous synthetic respondent built on an LLM produced coherent survey answers and escaped most current detection methods. The technique that promises to replace the participant thus threatens the data of real participants. Some of the authors of UXAgent, a system discussed in Section~\ref{sec:approaches}, work at Amazon, the company that runs Mechanical Turk (LU \textit{et al.}, 2025).

\subsection{Conflicts with academic research}

Academic research and commodity production ask different questions of users. A firm wants to know whether a product works well enough to sell. Research wants to know how people live with technology and what that experience teaches. \reading{In this paper's reading, the simulated user serves the first question better than the second.}

The conflict appears in at least four places.

The first is the value of knowledge. Agnew \textit{et al.} (2024) argue that substitution proposals conflict with the values that ground research with human participants: representation, inclusion and understanding. Those values do not enter the cost calculation of a product cycle.

The second is the legitimacy of methods. Kapania \textit{et al.} (2025) report that qualitative researchers fear the delegitimisation of their methods. If a synthetic interview costs cents and answers instantly, an interview with a person starts to look like an unjustified luxury.

The third is the university itself. \reading{This is again the paper's reading.} Pressure for academic productivity, measured in number of papers, pushes researchers toward the same shortcuts that industry seeks. Much of the literature on synthetic users comes from laboratories tied to large technology firms such as Amazon and Google (LU \textit{et al.}, 2025; DUAN \textit{et al.}, 2024a). The boundary between research and product development becomes less clear.

The fourth is research ethics. In Brazil, Resolution 510/2016 of the National Health Council (CNS) protects human participants in the humanities and social sciences (BRASIL, 2016). A study done only with synthetic users has no human participants in the sense of that resolution. It escapes protection, even though it produces decisions that affect people. The Brazilian HCI community has named the ethics of human participation in research as one of its grand challenges for 2025--2035 (PEREIRA; DARIN; SILVEIRA, 2024). Section~\ref{sec:ethics} returns to this point.

\subsection{Participatory Design as a limit case}\label{sec:pd}

Participatory Design makes the conflict sharper than any other method. It emerged in Scandinavia in the 1970s, in projects with trade unions that discussed the arrival of computers in factories and offices. Ehn (1992) describes that origin as a struggle for democracy at work. Labour laws of the period, such as the Swedish Joint Regulation Act, created conditions for unions to influence the systems that would change their members' work.

Kensing and Blomberg (1998) organise the field around three issues: the politics of design, the nature of participation, and the methods, tools and techniques that make participation possible. The first issue shapes the other two. In Participatory Design, people do not take part in order to supply data to the designer. They take part because they have a right to influence a technology that will change their lives.

Latin America has its own root for this idea. Participatory action research, systematised by Fals Borda (1987), treats research as a joint production of knowledge with communities, oriented toward social change. Escobar (2018) extends the argument to design for autonomy, in which communities design the conditions of their own lives. Lima and Lima (2025) recently proposed a context-sensitive Participatory Design for Brazil, based on third-wave HCI and decolonial thought.

This premise makes the simulated participant a contradiction in terms. An agent can answer questions. It cannot make claims, negotiate, refuse or organise. It has no interests to defend. \reading{In Participatory Design, simulating the participant removes the political subject of the method and keeps only its appearance.}

Ehn (1992) adds an epistemological argument. Drawing on Wittgenstein, he holds that practical skill is tacit and cannot be fully formalised. For that reason, Participatory Design works with prototypes, mock-ups and shared experience, so that designers and users learn from each other. A simulated user only knows what someone has already written down. Tacit knowledge, by definition, is what has not been written down.

\reading{The same reasoning applies, in different degrees, to other methods that depend on the person's presence.} Ethnography depends on the researcher being where the technology is used. Qualitative interviewing depends on the relationship between the one who asks and the one who answers (KAPANIA \textit{et al.}, 2025). Accessibility testing depends on people with disabilities, who sit at the edges of the distributions that models tend to erase (WANG; MORGENSTERN; DICKERSON, 2025). The more a method treats the participant as a subject rather than a data source, the less room it leaves for simulation. Participatory Design sits at the end of that scale. There, generative AI can help prepare materials or rehearse facilitation. It cannot take the participant's seat.

\section{A genealogy of the simulated user}\label{sec:genealogy}

\subsection{The idealised user: cognitive models}

The first form of the simulated user appeared at the Xerox Palo Alto Research Center (PARC). In 1983, Card, Moran and Newell published \textit{The Psychology of Human-Computer Interaction}, which proposed an applied psychology of the users of interactive systems (CARD; MORAN; NEWELL, 1983). The book brought two instruments. The GOMS model describes a task in goals, operators, methods and selection rules. The keystroke-level model (KLM) assigns a time to each elementary operation and adds them up to predict the duration of a task.

The user of these models is idealised. The KLM predicts the time of an expert who performs a routine task without errors. Learning, error and fatigue are left out. An explicit theory of the mind made up for the absence of the person.

Two decades later, John \textit{et al.} (2004) observed that such models had little impact in practice because they were hard to build. They created CogTool. In it, a designer demonstrates a task on an HTML mock-up and the system generates the model automatically. The model runs in the ACT-R cognitive architecture, which simulates perception and movement, and it interacts with the mock-up. CogTool is the most direct technical antecedent of current agents: a simulated user that operates an interface and produces performance data.

\subsection{The interpreted user: expert inspection}

The second form comes from low-cost usability engineering. Nielsen and Molich (1990) introduced heuristic evaluation, in which evaluators examine an interface against a short list of principles. They showed that a single evaluator found between 20\% and 51\% of the problems, while the combined work of three to five evaluators did well. In the same paper, they judged automatic evaluation infeasible except for narrow checks.

Heuristic evaluation replaces the user with an interpreter of the user. Generative AI, as Section~\ref{sec:approaches} shows, automates that interpreter.

Ivory and Hearst (2001) organised evaluation methods in a taxonomy centred on automation. They saw automation as a way to extend existing methods, justified by the cost of testing with people. The cost argument runs through this whole genealogy.

\subsection{The founding conflict: situated action}

The same laboratory that produced GOMS produced its strongest critique. In \textit{Plans and Situated Actions}, Suchman (1987) studied people using a photocopier with a help system designed on the planning model of cognitive science. The interaction rarely followed the plan, neither the person's nor the machine's. For Suchman, plans are vague resources for dealing with the contingencies of a situation, not programs that determine action. The problem of cognitive science lay in substituting the plan for the action and the representation of the circumstance for the circumstance itself.

That critique reaches every form of the simulated user. A GOMS model is a formalised plan. A generative agent also works through planning, memory and reflection. In both cases, the situation of use enters the test only as a representation.

\subsection{The statistical user: the data turn}

Between cognitive models and LLMs there is an intermediate stage. The user model stops coming from a theory and starts coming from data on human behaviour. Swearngin and Li (2019) collected crowdsourced labels on which elements of mobile interfaces people perceived as tappable. With those labels they trained a neural network that predicts that perception, with mean precision of 90.2\% and recall of 87.0\%. Their tool flags mismatches between what the designer intended and what users will probably perceive.

The user of this stage is an average of thousands of real people. The simulated task is narrow and measurable, and validation compares the model with people. In Brazil, Iannuzzi \textit{et al.} (2025) mapped this line of work. Of 238 studies found, they analysed 47. The most frequent methods apply supervised models to interaction metrics, think-aloud verbalisations and interface images. The authors point to limits in generalising the models and in integrating them with HCI and UX specialists.

\subsection{The believable user: generative simulacra}

The current form of the simulated user got its name in two papers from Joon Sung Park's group at Stanford. In \textit{Social Simulacra}, Park \textit{et al.} (2022) generated thousands of members of an online community and their interactions from a designer's description of goals, rules and personas. The technique relied on an explicit observation: the model's training data already contain a wide range of social media behaviour. In their evaluations, participants were often unable to tell simulated interactions from real ones. In \textit{Generative Agents}, Park \textit{et al.} (2023) created agents with memory, reflection and planning, who lived in a virtual town of 25 inhabitants. The declared evaluation criterion was the believability of their behaviour.

The word \textit{simulacrum} defines the concept. The agent does not aim to predict what a person would do. It aims to produce behaviour that looks human.

\subsection{A revisitation: computational rationality}

Not all recent work follows the generative path. Oulasvirta, Jokinen and Howes (2022) explicitly return to the question asked by Card, Moran and Newell in 1983 about how people interact with computers. They reframe it as a question of adaptation: users act as well as they can within the limits of their cognition, the device and the environment. This theory, called computational rationality, can be expressed in models that explain and predict interaction.

\reading{In this paper's reading, the revisitation works as a silent counterpoint to LLM-based simulation.} It keeps the project of simulating the user but demands an explicit theory of behaviour. A generative agent simulates without such a theory. Its behaviour comes from the statistics of text, not from a hypothesis about the mind. The genealogy splits here into one branch that models cognition and another that models language about cognition.

\subsection{The shift in the validity criterion}\label{sec:shift}

\reading{The genealogy reveals a change in what counts as a good user model. This is the paper's central analytical claim.} The models of Card, Moran and Newell (1983) and of John \textit{et al.} (2004) were judged by precision: the predicted time had to match the measured time. The model of Swearngin and Li (2019) was also judged by agreement with people. The simulacra of Park \textit{et al.} (2022; 2023) replaced precision with believability.

That change has consequences. An idealised but precise user says something verifiable about a narrow type of task. A believable but imprecise user says something plausible about any task. The second covers a much larger territory and is much harder to refute. Table~\ref{tab:genealogy} summarises the sequence.

\begin{table}[ht]
\centering\small
\caption{Forms of the simulated user in HCI.}\label{tab:genealogy}
\begin{tabularx}{\textwidth}{@{}p{2.1cm}LLLL@{}}
\toprule
\textbf{Period} & \textbf{Figure of the user} & \textbf{Source of the model} & \textbf{Validity criterion} & \textbf{Key works} \\
\midrule
1980s--2000s & Idealised expert, without errors & Theory of the mind (GOMS, KLM, ACT-R) & Precision of predicted time & Card, Moran and Newell (1983); John \textit{et al.} (2004) \\
1990s--2000s & Interpreted by an expert & Design principles applied by evaluators & Number of problems found & Nielsen and Molich (1990); Ivory and Hearst (2001) \\
2010s & Statistical average of real people & Behavioural data and machine learning & Agreement with human labels & Swearngin and Li (2019); Iannuzzi \textit{et al.} (2025) \\
2022 onward & Believable simulacrum & Statistics of text in LLMs & Believability to observers & Park \textit{et al.} (2022; 2023); Lu \textit{et al.} (2025) \\
2022 onward & Adaptive agent & Theory of bounded adaptation & Explanation and prediction & Oulasvirta, Jokinen and Howes (2022) \\
\bottomrule
\end{tabularx}
\end{table}

\section{Current approaches and their conflicts}\label{sec:approaches}

\subsection{The model as evaluator}

In the first approach, the LLM takes the place of the expert in heuristic evaluation. Duan \textit{et al.} (2024a) tested GPT-4 in a Figma plug-in that receives a screen and a set of guidelines and returns suggestions. They evaluated 51 interfaces with three sets of guidelines, compared the results with those of human experts and studied the tool with 12 expert designers. The model caught subtle errors and improved interface text, but its suggestions became less useful over iterations. The plug-in read the screen as a JSON data structure, one static screen at a time. In later work, the same group acknowledged that LLM-based techniques do not yet match human evaluators. Performance improved by 55\% when the model received interface critiques written by people (DUAN \textit{et al.}, 2024b).

Brazilian studies point the same way. Lima \textit{et al.} (2025) compared four generative AI systems in evaluating the websites of COP30 and COP29. The systems found problems of navigation, accessibility and terminology, but showed limits in cultural sensitivity and technical accessibility. In an extended version, Lima \textit{et al.} (2026) compared the systems' evaluations with the perceptions of ten human participants. People reported high satisfaction with the COP30 website, while the AI systems penalised it for structural deficiencies. Both groups agreed on the weakest criterion. The authors argue for hybrid models in which AI supplies systematic heuristic data and people add practical and sociocultural interpretation.

Matthaus \textit{et al.} (2026) studied the same approach in education. In an experiment with 24 software engineering students, they found no statistically significant difference between groups with and without generative AI, but the AI group had a lower mean number of correct identifications. Interaction logs showed mostly passive use, uncritical acceptance of answers, hallucinations, sycophantic responses and difficulty in noticing the tool's errors. The authors warn of cognitive delegation in conceptual learning.

\subsection{The model as participant}

In the second approach, the LLM takes the place of the participant. Hämäläinen, Tavast and Kunnari (2023) used GPT-3 to generate open-ended questionnaire answers about experiencing video games as art. Readers found it hard to tell synthetic answers from real ones. The authors recommended using such data for ideation and pilots, always followed by validation with real data.

In the social sciences, Argyle \textit{et al.} (2023) proposed the concept of algorithmic fidelity. On this view, a model's bias is fine-grained and linked to demographic groups, so conditioning the model on life stories reproduces the answer distributions of human subgroups. This is the strongest case for substitution.

The UXAgent system brings that logic to usability testing (LU \textit{et al.}, 2025). It generates thousands of personas and sends each one to navigate the website under evaluation, recording actions, video and an interview with the agent. The authors define its use as a rehearsal of the study protocol before the study with people. The five UX researchers who assessed the system praised it and also voiced concern about the future of UX research done with agents.

\subsection{Epistemological conflicts}\label{sec:epistemic}

An epistemological conflict, in this paper, is a tension about the validity of the knowledge produced. The question is whether what simulation produces can count as evidence about people.

\paragraph{Believability is not precision.} The group behind UXAgent measured whether agents reproduce real actions in 31,865 online shopping sessions with 230,965 user actions. The models predicted 11.86\% of human actions correctly (LU \textit{et al.}, 2026). Fine-tuning on real click data improved the result. The study puts precision back as the criterion and, in the terms of this genealogy, returns to the standard of the cognitive models. The Brazilian comparison by Lima \textit{et al.} (2026) shows the same gap in the evaluator approach: the AI's judgement and the users' experience diverged.

\paragraph{The average erases variation.} Bisbee \textit{et al.} (2024) found that synthetic means came close to real means in opinion surveys, but synthetic answers varied less than human ones, and regression coefficients diverged. The study contests algorithmic fidelity. Wang, Morgenstern and Dickerson (2025) tested four LLMs against 3,200 participants from 16 demographic identities. They found that LLMs misportray and flatten the groups they simulate, and that identity prompts tend to essentialise identities. They explain flattening by two causes: training rewards the most probable output, and scraped text rarely carries the identity of its author. In usability work, problems often appear at the edges of the distribution, such as an older person, a person with low vision or a person with little digital experience. \reading{A method that compresses variation tends to hide those cases.}

\paragraph{The situation is missing.} Kapania \textit{et al.} (2025) interviewed 19 qualitative researchers. Synthetic answers lacked palpability and contextual depth. The authors name the phenomenon the surrogate effect. \reading{This critique repeats Suchman's (1987) argument almost four decades later.} Usability is assessed in the encounter between a situated body and an artefact. An agent has no body, fatigue, haste or history with technology. It reads an interface as structured text.

\paragraph{Results do not repeat.} Bisbee \textit{et al.} (2024) showed that small changes in prompt wording changed the answers, and that the same prompt produced different results over three months. Providers update models without notice.

\paragraph{Real data become contaminated.} Westwood (2025) showed that coherence no longer indicates human authorship in online surveys. Remote unmoderated usability tests, common in UX practice, fall within the same zone of risk. In a preprint, Rothschild \textit{et al.} (2025) replied that the evidence does not yet show contamination to be real and large, so the size of the threat remains under debate.

\paragraph{The test confirms what is already known.} \reading{This is a hypothesis of the paper.} The model learns from texts that describe how people use interfaces, including guidelines and best practices. When it simulates a user, it tends to return that already written knowledge. The decline in usefulness across iterations reported by Duan \textit{et al.} (2024a) is consistent with this hypothesis, although the authors do not frame it that way.

\subsection{Ethical conflicts}\label{sec:ethics}

\paragraph{Consent and agency.} Kapania \textit{et al.} (2025) report that LLMs foreclose participants' consent and agency. A human participant can refuse, interrupt and contest a question. An agent does none of that of its own accord. Resolution 510/2016 defines free and informed consent as the participant's agreement ``livre de simulação, fraude, erro ou intimidação'', that is, free from simulation, fraud, error or intimidation (BRASIL, 2016, art. 2, V).

\paragraph{A regulatory gap.} The resolution applies to research that uses data obtained directly from participants, identifiable information or procedures with risks greater than those of everyday life (BRASIL, 2016, art. 1). A study done only with synthetic users fits none of those conditions. It tends to fall outside the system of Research Ethics Committees (CEP) and the National Research Ethics Commission (CONEP). \reading{In this paper's reading, the result is a study that speaks about people and guides product decisions that affect them without passing through any body that protects them.} The international literature reviewed does not discuss this gap, because it depends on the Brazilian regulatory framework.

\paragraph{The illusion of inclusion.} Agnew \textit{et al.} (2024) conclude that substitution proposals promise diversity without the presence of diverse people. Participation is also the means by which people influence the technologies they will use.

\paragraph{Harm to marginalised groups.} Wang, Morgenstern and Dickerson (2025) link flattening to a history of one-dimensional portrayals of marginalised groups. They urge caution whenever the participant's identity matters for the task. In accessibility testing, it almost always matters.

\paragraph{Training data and invisible labour.} As noted in Section~\ref{sec:problem}, the synthetic user carries the unconsented voice of real people and the work of data annotators (KAPANIA \textit{et al.}, 2025; GRAY; SURI, 2019).

\paragraph{The protection argument.} The ethical conflict has two sides. Lu \textit{et al.} (2025) argue that simulating before testing protects participants from badly designed protocols. The argument holds for pilots. It weakens when the pilot takes the place of the study, which is what the cost motivation tends to produce.

Table~\ref{tab:conflicts} summarises the conflicts by type of use.

\begin{table}[ht]
\centering\small
\caption{Conflicts by type of use of LLM-based simulation.}\label{tab:conflicts}
\begin{tabularx}{\textwidth}{@{}p{2.8cm}LLp{3.2cm}@{}}
\toprule
\textbf{Use} & \textbf{Epistemological conflict} & \textbf{Ethical conflict} & \textbf{Sources} \\
\midrule
Automated heuristic evaluation & Below human performance; screen read as data; possible circularity & Low with human review; risk of cognitive delegation in education & Duan \textit{et al.} (2024a; 2024b); Lima \textit{et al.} (2025; 2026); Matthaus \textit{et al.} (2026) \\
Pilot of a study protocol & Low, if findings do not count as evidence about users & May protect real participants & Lu \textit{et al.} (2025); Hämäläinen, Tavast and Kunnari (2023) \\
Testing with synthetic users & Believability without precision; flattened variation; low reproducibility & No consent; outside the CEP/CONEP system & Lu \textit{et al.} (2026); Bisbee \textit{et al.} (2024); Brasil (2016) \\
Personas of specific groups & Portrayal from outside the group; essentialised identity & Representational harm & Wang, Morgenstern and Dickerson (2025) \\
Synthetic interviews & No palpability or context & Agency foreclosed; unconsented training data & Kapania \textit{et al.} (2025) \\
Broad substitution & Contested algorithmic fidelity & Illusion of inclusion & Argyle \textit{et al.} (2023); Agnew \textit{et al.} (2024) \\
Unmoderated online collection & Contamination of human data & Trust in the evidence base & Westwood (2025) \\
Participatory Design & Tacit knowledge out of reach & Political subject removed & Ehn (1992); Kensing and Blomberg (1998) \\
\bottomrule
\end{tabularx}
\end{table}

\section{The view from Latin America}\label{sec:latam}

\subsection{What the region produces}

The Latin American work found in this review concentrates on the evaluator approach. Brazilian studies compare generative AI with human evaluators and users (LIMA \textit{et al.}, 2025; 2026), map machine learning in usability evaluation (IANNUZZI \textit{et al.}, 2025) and study the effects of generative AI on learning heuristic evaluation (MATTHAUS \textit{et al.}, 2026). In Peru, Lecaros \textit{et al.} (2024) automated the Granollers heuristic evaluation method with a support system that does not rely on AI. The common pattern treats AI as support for human judgement and argues for hybrid arrangements.

\subsection{An absence}

The review found no academic study from Latin America, in Portuguese or Spanish, on synthetic users replacing participants in UX testing. Searches in Spanish returned blog posts and marketing material from companies that sell the service. \reading{In this paper's reading, the absence is itself a finding.} The tools arrive in the region as products before they arrive as objects of research. If models misportray groups from the outside (WANG; MORGENSTERN; DICKERSON, 2025), the risk grows for populations that are under-represented in training data. Ricaurte (2019) treats data-centred epistemologies as an expression of the coloniality of power. Faustino and Lippold (2023) analyse digital colonialism from a Fanonian perspective. Both frames suggest that a Latin American user produced by a model trained mostly on texts from the global North is not a neutral approximation.

\subsection{Resources from the region}

The region also offers resources to address the problem. The Brazilian HCI community defined seven grand challenges for 2025--2035. Three of them touch this paper directly: ethics and responsibility in human participation in HCI research (GC2), plurality and decoloniality in HCI (GC3), and the implications of artificial intelligence for HCI (GC6) (PEREIRA; DARIN; SILVEIRA, 2024). Martins, Lana and Villela (2025) responded to the AI challenge by studying how ethical values are communicated in the interfaces of generative AI systems.

Semiotic engineering, developed in Brazil, treats the interface as a message from designers to users (DE SOUZA, 2005). Its communicability evaluation method relies on observing users' interaction and identifying moments of communicative breakdown (PRATES; DE SOUZA; BARBOSA, 2000). \reading{In this paper's reading, a synthetic user can produce a predicted breakdown, but not a breakdown lived by someone who did not understand the designer's message.} The method therefore offers a principled reason to keep people in evaluation.

Álvaro Vieira Pinto (2005) criticised the idea of a ``technological age'' as ideology and treated technology as a human production situated in history. \reading{Read through his work, the synthetic user appears less as an inevitable stage of technical progress and more as a choice made under specific conditions of production.} Participatory action research (FALS BORDA, 1987) and design for autonomy (ESCOBAR, 2018), discussed in Section~\ref{sec:pd}, complete this set of regional resources.

\section{Discussion}\label{sec:discussion}

\subsection{Believability as a product}

\reading{This section develops the paper's interpretation.} The shift from precision to believability (Section~\ref{sec:shift}) and the pressure of production (Section~\ref{sec:problem}) converge. A believable answer arrives fast and costs little. A precise answer requires people, time and measurement. When the criterion of validity becomes believability, the user model becomes something that can be sold as a service, because believability is a property that the model itself can generate. Precision, in contrast, can only be established against people. The studies that return to precision (LU \textit{et al.}, 2026; BISBEE \textit{et al.}, 2024) and the Brazilian comparison between AI judgement and user experience (LIMA \textit{et al.}, 2026) show the cost of that substitution.

\subsection{Two approaches, two weights}

Automating the evaluator inherits the limits of heuristic evaluation, which was always a method without users. Automating the participant is a larger break, because it removes the source that usability testing exists to consult. The evaluator approach is not free of problems, however. Matthaus \textit{et al.} (2026) show that, in education, it can replace the learning of evaluation with the delegation of evaluation. \reading{In Braverman's terms, the first approach risks deskilling the evaluator, while the second removes the participant.}

\subsection{A reflexive note on this paper}\label{sec:reflexive}

As stated in Section~\ref{sec:ai-use}, this paper was produced with an LLM as research assistant. \reading{That process is itself an instance of the separation between conception and execution discussed in Section~\ref{sec:problem}.} The model performed tasks that a human research assistant would perform: searching, summarising, checking records, drafting and translating. It did not set the research question or the theoretical framing. The framing through the mode of production, the treatment of Participatory Design and the turn toward Latin American sources came from the author's requests at different points of the dialogue.

The process also produced errors that verification caught. An early draft stated that Amazon had funded UXAgent. The sources only support the claim that some authors work at Amazon, and the text was corrected. Access dates in an early version of the reference list were inconsistent and were corrected. These cases support the paper's argument in a modest way. Text produced by an LLM is believable before it is verified. Its precision had to be established by checking each claim against primary records, which is the same move that Lu \textit{et al.} (2026) make for simulated users.

The paper does not claim that this mode of production is neutral. It reports it so that readers can weigh the work accordingly.

\section{Conclusions}\label{sec:conclusions}

The first conclusion is that the simulated user is an old figure in HCI. It has changed form with the available technique. It was an idealised expert in the 1980s, an interpreter of the user in heuristic evaluation, a statistical average in the 2010s and a believable simulacrum from 2022 onward. Each form inherited the cost argument and Suchman's objection about the situation.

The second conclusion is that the decisive change was not technical. It was the replacement of precision by believability as the criterion of validity. That change lets the simulated user speak about any task, at the price of saying little that can be verified. Recent studies that return to precision find weak results.

The third conclusion is that the literature converges on a minimal consensus. Simulation can help rehearse protocols and triage problems early. It does not replace testing with people. The real disagreement concerns the nature of the limits. Technical work treats failures as gaps that more data and fine-tuning will close. Critical work treats them as properties of the method. A model trained to produce the most probable output cannot represent the improbable person, who is often the one who reveals the problem of an interface.

The fourth conclusion is that the two approaches weigh differently. Automating the evaluator risks deskilling. Automating the participant removes the source of knowledge that testing exists to reach. In Participatory Design, it removes the political subject of the method.

The fifth conclusion concerns Latin America. The region uses generative AI as support for inspection and has not yet studied synthetic users. That gap matters because the people most likely to be misportrayed by models trained elsewhere are also the people least present in the research on those models.

\section{Open problems and future work}\label{sec:future}

\paragraph{Ethics and regulation in Brazil.} It remains to be examined how the CEP/CONEP system should treat studies that use only synthetic users to decide about products used by people. One line of work is documentary: analysing opinions, resolutions and consultations on the topic. Another is propositional: formulating criteria for disclosure and review of such studies.

\paragraph{Populations of the global South.} No empirical study on simulated users with Brazilian or Latin American populations was found. A replication of Wang, Morgenstern and Dickerson (2025) with Brazilian groups, varying region, income and digital literacy, would test whether flattening is stronger outside the centres where training data are produced.

\paragraph{Body and situation.} The literature points to the lack of context and palpability but does not develop it as a phenomenological problem. Theoretical work can connect Suchman (1987) with phenomenology of the body to explain why the encounter between a body and an artefact does not reduce to text.

\paragraph{Other critical traditions.} Activity ergonomics separates prescribed tasks from real activity, and semiotic engineering treats the interface as communication. Both can offer counterpoints to the simulated user that this review only began to explore.

\paragraph{Mixed validation.} Protocols are needed that combine simulation and testing with people and measure what each finds. A comparative study could record which problems only agents detect, which only people detect and which both detect. Accessibility testing is a priority case, because it concentrates the users at the edges of the distribution.

\paragraph{Contamination of remote tests.} Westwood (2025) studied opinion surveys. The presence of agents in remote unmoderated usability tests, and methods to verify human authorship, remain to be measured.

\paragraph{Circularity.} The hypothesis that simulated users return the knowledge already written about usability can be tested. One path is to compare agent performance on conventional interfaces and on interfaces that break known patterns.

\paragraph{Reproducibility.} Standards are needed to record the model version, the prompt, the sampling settings and the date in studies with synthetic users, so that others can repeat them.

\section*{References}
\addcontentsline{toc}{section}{References}
\begin{abntrefs}

AGNEW, William \textit{et al.} The illusion of artificial inclusion. \textit{In}: CHI CONFERENCE ON HUMAN FACTORS IN COMPUTING SYSTEMS, 2024, Honolulu. \textbf{Proceedings} [\ldots]. New York: ACM, 2024. p. 1--12. DOI: \href{https://doi.org/10.1145/3613904.3642703}{10.1145/3613904.3642703}.

ARGYLE, Lisa P. \textit{et al.} Out of one, many: using language models to simulate human samples. \textbf{Political Analysis}, Cambridge, v. 31, n. 3, p. 337--351, 2023. DOI: \href{https://doi.org/10.1017/pan.2023.2}{10.1017/pan.2023.2}. Preprint: \url{https://arxiv.org/abs/2209.06899}.

BISBEE, James \textit{et al.} Synthetic replacements for human survey data? The perils of large language models. \textbf{Political Analysis}, Cambridge, v. 32, n. 4, p. 401--416, 2024. DOI: \href{https://doi.org/10.1017/pan.2024.5}{10.1017/pan.2024.5}.

BRASIL. Conselho Nacional de Saúde. \textbf{Resolução nº 510, de 7 de abril de 2016}. Dispõe sobre as normas aplicáveis a pesquisas em Ciências Humanas e Sociais. \textbf{Diário Oficial da União}: seção 1, Brasília, DF, n. 98, p. 44, 24 maio 2016. \refavail{} \url{http://bvsms.saude.gov.br/bvs/saudelegis/cns/2016/res0510_07_04_2016.html}. \refaccess{} 23 \refnov{} 2026.

BRAVERMAN, Harry. \textbf{Labor and monopoly capital}: the degradation of work in the twentieth century. New York: Monthly Review Press, 1974. \refreissue{} 1998. ISBN 9781583673751.

CARD, Stuart K.; MORAN, Thomas P.; NEWELL, Allen. \textbf{The psychology of human-computer interaction}. Hillsdale: Lawrence Erlbaum Associates, 1983. ISBN 0898592437. \refreissue{} Routledge, 2018. \refavail{} \url{https://www.routledge.com/products/9780203736166}. \refaccess{} 23 \refnov{} 2026.

DE SOUZA, Clarisse Sieckenius. \textbf{The semiotic engineering of human-computer interaction}. Cambridge, MA: MIT Press, 2005. (Acting with Technology). \refavail{} \url{https://mitpressbookstore.mit.edu/book/9780262527095}. \refaccess{} 23 \refnov{} 2026.

DUAN, Peitong \textit{et al.} Generating automatic feedback on UI mockups with large language models. \textit{In}: CHI CONFERENCE ON HUMAN FACTORS IN COMPUTING SYSTEMS, 2024, Honolulu. \textbf{Proceedings} [\ldots]. New York: ACM, 2024a. Preprint: \url{https://arxiv.org/abs/2403.13139}. DOI: \href{https://doi.org/10.1145/3613904.3642782}{10.1145/3613904.3642782}.

DUAN, Peitong \textit{et al.} UICrit: enhancing automated design evaluation with a UI critique dataset. \textit{In}: ANNUAL ACM SYMPOSIUM ON USER INTERFACE SOFTWARE AND TECHNOLOGY, 37., 2024, Pittsburgh. \textbf{Proceedings} [\ldots]. New York: ACM, 2024b. art. 46. DOI: \href{https://doi.org/10.1145/3654777.3676381}{10.1145/3654777.3676381}.

EHN, Pelle. Scandinavian design: on participation and skill. \textit{In}: ADLER, Paul S.; WINOGRAD, Terry A. (ed.). \textbf{Usability}: turning technologies into tools. New York: Oxford University Press, 1992. p. 96--132. ISBN 9780195075106. \refavail{} \url{https://dblp.org/db/books/collections/AW1992.html}. \refaccess{} 23 \refnov{} 2026.

ESCOBAR, Arturo. \textbf{Designs for the pluriverse}: radical interdependence, autonomy, and the making of worlds. Durham: Duke University Press, 2018. DOI: \href{https://doi.org/10.1215/9780822371816}{10.1215/9780822371816}.

FALS BORDA, Orlando. The application of participatory action-research in Latin America. \textbf{International Sociology}, London, v. 2, n. 4, p. 329--347, 1987. DOI: \href{https://doi.org/10.1177/026858098700200401}{10.1177/026858098700200401}.

FAUSTINO, Deivison; LIPPOLD, Walter. \textbf{Colonialismo digital}: por uma crítica hacker-fanoniana. São Paulo: Boitempo, 2023. 208 p. ISBN 9786557172254.

GRAY, Mary L.; SURI, Siddharth. \textbf{Ghost work}: how to stop Silicon Valley from building a new global underclass. Boston: Houghton Mifflin Harcourt, 2019. ISBN 9781328566249.

HÄMÄLÄINEN, Perttu; TAVAST, Mikke; KUNNARI, Anton. Evaluating large language models in generating synthetic HCI research data: a case study. \textit{In}: CHI CONFERENCE ON HUMAN FACTORS IN COMPUTING SYSTEMS, 2023, Hamburg. \textbf{Proceedings} [\ldots]. New York: ACM, 2023. art. 433, p. 1--19. DOI: \href{https://doi.org/10.1145/3544548.3580688}{10.1145/3544548.3580688}.

IANNUZZI, João Carlos Guimarães \textit{et al.} Integração de machine learning na avaliação de usabilidade: avanços, desafios e aplicações futuras. \textit{In}: SIMPÓSIO BRASILEIRO SOBRE FATORES HUMANOS EM SISTEMAS COMPUTACIONAIS (IHC), 24., 2025, Belo Horizonte. \textbf{Anais} [\ldots]. Porto Alegre: Sociedade Brasileira de Computação, 2025. p. 1513--1537. DOI: \href{https://doi.org/10.5753/ihc.2025.10959}{10.5753/ihc.2025.10959}.

IVORY, Melody Y.; HEARST, Marti A. The state of the art in automating usability evaluation of user interfaces. \textbf{ACM Computing Surveys}, New York, v. 33, n. 4, p. 470--516, \refdec{} 2001. DOI: \href{https://doi.org/10.1145/503112.503114}{10.1145/503112.503114}.

JOHN, Bonnie E. \textit{et al.} Predictive human performance modeling made easy. \textit{In}: SIGCHI CONFERENCE ON HUMAN FACTORS IN COMPUTING SYSTEMS, 2004, Vienna. \textbf{Proceedings} [\ldots]. New York: ACM, 2004. p. 455--462. DOI: \href{https://doi.org/10.1145/985692.985750}{10.1145/985692.985750}.

KAPANIA, Shivani \textit{et al.} `Simulacrum of stories': examining large language models as qualitative research participants. \textit{In}: CHI CONFERENCE ON HUMAN FACTORS IN COMPUTING SYSTEMS, 2025, Yokohama. \textbf{Proceedings} [\ldots]. New York: ACM, 2025. art. 489, p. 1--17. DOI: \href{https://doi.org/10.1145/3706598.3713220}{10.1145/3706598.3713220}. Preprint: \url{https://arxiv.org/abs/2409.19430}.

KENSING, Finn; BLOMBERG, Jeanette. Participatory design: issues and concerns. \textbf{Computer Supported Cooperative Work}, Dordrecht, v. 7, p. 167--185, 1998. DOI: \href{https://doi.org/10.1023/A:1008689307411}{10.1023/A:1008689307411}.

LECAROS, Adrian \textit{et al.} Automation of Granollers heuristic evaluation method using a developed support system. \textit{In}: IBEROAMERICAN WORKSHOP ON HUMAN-COMPUTER INTERACTION (HCI-COLLAB), 9., 2023, Buenos Aires. \textbf{Revised selected papers}. Cham: Springer, 2024. p. 93--108. DOI: \href{https://doi.org/10.1007/978-3-031-57982-0_8}{10.1007/978-3-031-57982-0\_8}.

LIMA, Danilo Teixeira \textit{et al.} Can AI judge usability? A comparative analysis of generative tools on climate conference websites. \textit{In}: SIMPÓSIO BRASILEIRO SOBRE FATORES HUMANOS EM SISTEMAS COMPUTACIONAIS (IHC), 24., 2025, Belo Horizonte. \textbf{Anais} [\ldots]. Porto Alegre: Sociedade Brasileira de Computação, 2025. p. 497--517. DOI: \href{https://doi.org/10.5753/ihc.2025.10805}{10.5753/ihc.2025.10805}.

LIMA, Danilo T. \textit{et al.} Human perceptions meet AI analysis: a usability evaluation of climate conference websites. \textbf{Journal on Interactive Systems}, Porto Alegre, v. 17, n. 1, p. 1001--1022, 2026. DOI: \href{https://doi.org/10.5753/jis.2026.7129}{10.5753/jis.2026.7129}.

LIMA, Júlia G.; LIMA, Tamara C. Alves de. Blueprint para design participativo sensível ao contexto: resposta aos desafios GC3 e GC4 do GrandIHC-BR. \textit{In}: SIMPÓSIO BRASILEIRO SOBRE FATORES HUMANOS EM SISTEMAS COMPUTACIONAIS (IHC), 24., 2025, Belo Horizonte. \textbf{Anais Estendidos} [\ldots]. Porto Alegre: Sociedade Brasileira de Computação, 2025. p. 391--396. DOI: \href{https://doi.org/10.5753/ihc_estendido.2025.16215}{10.5753/ihc\_estendido.2025.16215}.

LU, Yuxuan \textit{et al.} UXAgent: an LLM agent-based usability testing framework for web design. \textit{In}: CHI CONFERENCE ON HUMAN FACTORS IN COMPUTING SYSTEMS, 2025, Yokohama. \textbf{Extended Abstracts} [\ldots]. New York: ACM, 2025. DOI: \href{https://doi.org/10.1145/3706599.3719729}{10.1145/3706599.3719729}.

LU, Yuxuan \textit{et al.} Can LLM agents simulate multi-turn human behavior? Evidence from real online customer behavior data. \textit{In}: ANNUAL MEETING OF THE ASSOCIATION FOR COMPUTATIONAL LINGUISTICS, 64., 2026, San Diego. \textbf{Proceedings} [\ldots]: volume 1: long papers. [\textit{S. l.}]: ACL, 2026. p. 43961--43977. \refavail{} \url{https://aclanthology.org/2026.acl-long.2034/}. \refaccess{} 23 \refnov{} 2026.

MARTINS, Helena Cristo; LANA, Cristiane Aparecida; VILLELA, Maria Lúcia Bento. Do diagnóstico à ação: avanços na pesquisa sobre ética em IA aplicada à IHC. \textit{In}: SIMPÓSIO BRASILEIRO SOBRE FATORES HUMANOS EM SISTEMAS COMPUTACIONAIS (IHC), 24., 2025, Belo Horizonte. \textbf{Anais Estendidos} [\ldots]. Porto Alegre: Sociedade Brasileira de Computação, 2025. p. 386--390. DOI: \href{https://doi.org/10.5753/ihc_estendido.2025.16210}{10.5753/ihc\_estendido.2025.16210}.

MATTHAUS, Fabiann \textit{et al.} Aprender ou delegar? Um estudo sobre pensamento crítico no uso de IA generativa em avaliação heurística. \textit{In}: SIMPÓSIO BRASILEIRO DE INFORMÁTICA NA EDUCAÇÃO (SBIE), 37., 2026, Goiânia. \textbf{Anais} [\ldots]. Porto Alegre: Sociedade Brasileira de Computação, 2026. p. 1077--1091. DOI: \href{https://doi.org/10.5753/sbie.2026.27476}{10.5753/sbie.2026.27476}.

NIELSEN, Jakob; MOLICH, Rolf. Heuristic evaluation of user interfaces. \textit{In}: SIGCHI CONFERENCE ON HUMAN FACTORS IN COMPUTING SYSTEMS, 1990, Seattle. \textbf{Proceedings} [\ldots]. New York: ACM, 1990. p. 249--256. DOI: \href{https://doi.org/10.1145/97243.97281}{10.1145/97243.97281}.

OULASVIRTA, Antti; JOKINEN, Jussi P. P.; HOWES, Andrew. Computational rationality as a theory of interaction. \textit{In}: CHI CONFERENCE ON HUMAN FACTORS IN COMPUTING SYSTEMS, 2022, New Orleans. \textbf{Proceedings} [\ldots]. New York: ACM, 2022. art. 359, p. 1--14. DOI: \href{https://doi.org/10.1145/3491102.3517739}{10.1145/3491102.3517739}.

PARK, Joon Sung \textit{et al.} Social simulacra: creating populated prototypes for social computing systems. \textit{In}: ANNUAL ACM SYMPOSIUM ON USER INTERFACE SOFTWARE AND TECHNOLOGY, 35., 2022, Bend. \textbf{Proceedings} [\ldots]. New York: ACM, 2022. DOI: \href{https://doi.org/10.1145/3526113.3545616}{10.1145/3526113.3545616}.

PARK, Joon Sung \textit{et al.} Generative agents: interactive simulacra of human behavior. \textit{In}: ANNUAL ACM SYMPOSIUM ON USER INTERFACE SOFTWARE AND TECHNOLOGY, 36., 2023, San Francisco. \textbf{Proceedings} [\ldots]. New York: ACM, 2023. DOI: \href{https://doi.org/10.1145/3586183.3606763}{10.1145/3586183.3606763}.

PEREIRA, Roberto; DARIN, Ticianne; SILVEIRA, Milene Selbach. GranDIHC-BR: grand research challenges in human-computer interaction in Brazil for 2025--2035. \textit{In}: SIMPÓSIO BRASILEIRO SOBRE FATORES HUMANOS EM SISTEMAS COMPUTACIONAIS (IHC), 23., 2024, Brasília. \textbf{Anais} [\ldots]. Porto Alegre: Sociedade Brasileira de Computação, 2024. p. 915--938. \refavail{} \url{https://sol.sbc.org.br/index.php/ihc/article/view/32921}. \refaccess{} 23 \refnov{} 2026.

PINTO, Álvaro Vieira. \textbf{O conceito de tecnologia}. Rio de Janeiro: Contraponto, 2005. 2 v. \refcatalog{} \url{https://busca.acervo.nic.br/vufind/Record/2073}. \refaccess{} 23 \refnov{} 2026.

PRATES, Raquel O.; DE SOUZA, Clarisse S.; BARBOSA, Simone D. J. A method for evaluating the communicability of user interfaces. \textbf{interactions}, New York, v. 7, n. 1, p. 31--38, 2000. DOI: \href{https://doi.org/10.1145/328595.328608}{10.1145/328595.328608}.

QIN, Yigang; CHEON, EunJeong. Labor, capital, and machine: toward a labor process theory for HCI. \textit{In}: CHI CONFERENCE ON HUMAN FACTORS IN COMPUTING SYSTEMS, 2026, Barcelona. \textbf{Proceedings} [\ldots]. New York: ACM, 2026. 20 p. DOI: \href{https://doi.org/10.1145/3772318.3791147}{10.1145/3772318.3791147}. Preprint: \url{https://arxiv.org/abs/2602.10548}.

RICAURTE, Paola. Data epistemologies, the coloniality of power, and resistance. \textbf{Television \& New Media}, Thousand Oaks, v. 20, n. 4, p. 350--365, 2019. DOI: \href{https://doi.org/10.1177/1527476419831640}{10.1177/1527476419831640}.

ROTHSCHILD, David M.; BARARI, Soubhik; BUSKIRK, Trent D.; GORDON, Andrew; HILLYGUS, D. Sunshine. Reply to Westwood: questioning the empirical evidence that AI survey contamination is real and substantial. \textbf{SocArXiv}, 2025. Preprint. DOI: \href{https://doi.org/10.31235/osf.io/ykxwj}{10.31235/osf.io/ykxwj}.

SUCHMAN, Lucy A. \textbf{Plans and situated actions}: the problem of human-machine communication. Cambridge: Cambridge University Press, 1987. 2. ed.: \textit{Human-machine reconfigurations}, 2007. DOI: \href{https://doi.org/10.1017/CBO9780511808418}{10.1017/CBO9780511808418}.

SWEARNGIN, Amanda; LI, Yang. Modeling mobile interface tappability using crowdsourcing and deep learning. \textit{In}: CHI CONFERENCE ON HUMAN FACTORS IN COMPUTING SYSTEMS, 2019, Glasgow. \textbf{Proceedings} [\ldots]. New York: ACM, 2019. DOI: \href{https://doi.org/10.1145/3290605.3300305}{10.1145/3290605.3300305}.

WANG, Angelina; MORGENSTERN, Jamie; DICKERSON, John P. Large language models that replace human participants can harmfully misportray and flatten identity groups. \textbf{Nature Machine Intelligence}, London, v. 7, n. 3, p. 400--411, 2025. DOI: \href{https://doi.org/10.1038/s42256-025-00986-z}{10.1038/s42256-025-00986-z}.

WESTWOOD, Sean J. The potential existential threat of large language models to online survey research. \textbf{Proceedings of the National Academy of Sciences}, Washington, DC, v. 122, n. 47, e2518075122, 2025. DOI: \href{https://doi.org/10.1073/pnas.2518075122}{10.1073/pnas.2518075122}.

\end{abntrefs}

\clearpage
\selectlanguage{brazilian}
\renewcommand{\refavail}{Disponível em:}
\renewcommand{\refaccess}{Acesso em:}
\renewcommand{\refoct}{out.}
\renewcommand{\refdec}{dez.}
\renewcommand{\refnov}{nov.}
\renewcommand{\refreissue}{Reedição:}
\renewcommand{\refcatalog}{Registro em catálogo:}
\setcounter{section}{0}
\setcounter{table}{0}

\part*{Parte II: versão original em português}\label{pt-start}
\addcontentsline{toc}{part}{Parte II: versão original em português}

\begin{center}
{\Large\bfseries O usuário que não está lá:}\\[0.3em]
{\large genealogia e crítica dos usuários simulados por IA generativa em testes de UX}\\[1em]
José Carlos Porto Arcoverde Jr. (Mabuse)\\
\small Programa de Pós-Graduação em Design, Universidade Federal de Pernambuco (PPGDesign-UFPE)\\
\small CESAR School, Recife, Brasil
\end{center}

\begin{abstract}
\noindent
O teste de usabilidade depende de pessoas, e o tempo das pessoas não acelera no ritmo que a produção de software exige. Os modelos de linguagem de grande escala (LLMs) prometem agora retirar as pessoas do teste, gerando usuários simulados que navegam em sites, respondem questionários e dão entrevistas. Este artigo examina como a pesquisa acadêmica compreende essa promessa. Ele combina uma revisão crítica de literatura com uma leitura genealógica, que acompanha o usuário simulado desde os modelos cognitivos dos anos 1980 até os agentes generativos atuais. O artigo defende quatro teses. A primeira é que testar sem o usuário é um objetivo antigo da Interação Humano-Computador (IHC), e o custo sempre o justificou. A segunda é que o critério de validade dos modelos de usuário mudou da precisão preditiva para a credibilidade, e os estudos recentes que voltam à precisão relatam resultados fracos. A terceira é que os participantes simulados entram em conflito com os valores da pesquisa acadêmica e com métodos que tratam a participação como direito político, sendo o Design Participativo o caso-limite. A quarta é que a literatura latino-americana encontrada usa a IA generativa como apoio à inspeção por especialistas e não traz nenhum estudo sobre usuários sintéticos. Essa lacuna deixa em aberto perguntas sobre quem é representado e sobre a regulação da ética em pesquisa no Brasil. O artigo termina com uma agenda de problemas em aberto. Como parte do método, informa que a própria revisão usou um LLM como assistente de pesquisa e discute o que isso implica.

\medskip
\noindent\textbf{Palavras-chave:} teste de usabilidade; usuários sintéticos; modelos de linguagem; Design Participativo; processo de trabalho; ética em pesquisa; América Latina.
\end{abstract}

\begin{aidisclaimer}{Declaração de uso de IA generativa}
Este artigo foi preparado com o apoio de um modelo de linguagem de grande escala (Claude, da Anthropic, nas versões disponíveis entre outubro e novembro de 2026), usado como assistente de pesquisa. O modelo buscou a literatura, resumiu fontes, conferiu registros bibliográficos, redigiu o texto em português e o traduziu para o inglês. As perguntas de pesquisa, o enquadramento teórico e as decisões editoriais são do autor. Cada referência foi conferida em pelo menos um registro primário, e a Seção~\ref{pt-metodo} informa os limites dessa conferência. O modelo não é autor deste artigo. O autor assume total responsabilidade pelo conteúdo, inclusive por eventuais erros. As Seções~\ref{pt-uso-ia} e~\ref{pt-reflexiva} descrevem e discutem esse processo.
\end{aidisclaimer}

\section{Introdução}\label{pt-intro}

O teste de usabilidade consulta pessoas. Ele coloca alguém diante de uma interface, pede uma tarefa e observa o que acontece. Essa consulta leva tempo. É preciso recrutar participantes, agendar sessões, conduzir e analisar. Por isso, a IHC busca há muito tempo formas de avaliar interfaces sem usuários.

A IA generativa oferece a versão mais radical dessa busca até hoje. Um modelo de linguagem de grande escala (LLM) pode gerar milhares de usuários simulados. Cada um tem uma persona, uma memória e uma forma de falar. Esses agentes navegam em sites, respondem questionários e dão entrevistas. A literatura os chama de usuários sintéticos, amostras de silício ou agentes simuladores (ARGYLE \textit{et al.}, 2023; LU \textit{et al.}, 2025).

A promessa levanta três perguntas. A primeira é epistemológica: o que um usuário simulado permite saber sobre pessoas reais? A segunda é ética: o que se perde, e quem se prejudica, quando a pessoa sai da pesquisa? A terceira é política: a quem serve a aceleração que a simulação oferece?

Este artigo trata dessas perguntas por meio de uma revisão crítica de literatura organizada como genealogia. Suas contribuições são as seguintes:
\begin{itemize}
  \item uma genealogia do usuário simulado na IHC, dos modelos cognitivos aos agentes generativos, que identifica uma mudança no critério de validade dos modelos de usuário (Seção~\ref{pt-genealogia});
  \item uma leitura dos usuários sintéticos como resposta à pressão de tempo da produção capitalista de software, com base na teoria do processo de trabalho (Seção~\ref{pt-problema});
  \item uma análise dos conflitos que os participantes sintéticos criam com a pesquisa acadêmica e com o Design Participativo (Seções~\ref{pt-problema} e~\ref{pt-abordagens});
  \item um levantamento da produção latino-americana sobre o tema, que não encontrou estudos sobre usuários sintéticos (Seção~\ref{pt-latam});
  \item uma agenda de problemas em aberto (Seção~\ref{pt-futuro}).
\end{itemize}

A Seção~\ref{pt-metodo} descreve o método, incluindo o uso de um LLM como assistente de pesquisa. A Seção~\ref{pt-discussao} discute os resultados e a Seção~\ref{pt-conclusoes} conclui.

Ao longo do texto, as passagens que apresentam interpretação própria deste artigo, e não afirmação das fontes citadas, vêm marcadas como tal.

\section{Método}\label{pt-metodo}

\subsection{Abordagem}

O artigo combina dois procedimentos. O primeiro é uma revisão crítica da pesquisa sobre simulação de usuários e de avaliadores com LLMs em estudos de usabilidade e de interação. O segundo é uma leitura genealógica da ideia de usuário simulado. Essa leitura pergunta o que preparou a ideia, por afinidade ou por conflito. Em seguida, pergunta como a ideia foi formulada e como trabalhos posteriores a contestaram ou retomaram. Uma genealogia não pressupõe uma linha reta de progresso. Ela procura disputas, rupturas e retornos.

\subsection{Busca e seleção}

As buscas ocorreram em outubro de 2026. As informações sobre políticas da Seção~\ref{pt-uso-ia} foram atualizadas em novembro de 2026. As fontes incluíram a ACM Digital Library, o arXiv, a ACL Anthology, páginas de editoras (Cambridge University Press, SAGE, Springer, MIT Press), a SBC OpenLib, repositórios universitários e a base dblp. As consultas foram feitas em inglês, português e espanhol. Elas combinaram termos sobre usabilidade, experiência do usuário, avaliação heurística, usuários sintéticos, agentes LLM, participantes de pesquisa e personas com termos sobre Design Participativo, trabalho e colonialidade. As listas de referências dos trabalhos centrais foram seguidas para encontrar outros estudos.

Um trabalho foi incluído quando tratava de pelo menos um destes temas: simulação de usuários ou de avaliadores com LLMs; críticas à substituição de participantes humanos por IA; antecedentes históricos da modelagem de usuários e da avaliação automática; ou os referenciais teóricos usados na análise. O corpus final tem 42 trabalhos.

\subsection{Uso de IA generativa neste estudo}\label{pt-uso-ia}

Este artigo foi produzido por meio de um diálogo extenso entre o autor e um LLM (Claude, da Anthropic) usado como assistente de pesquisa. O modelo fez buscas na web, resumiu páginas de editoras e resumos, conferiu registros bibliográficos, redigiu o texto em português e produziu a versão em inglês. O autor definiu as perguntas de pesquisa e o enquadramento teórico. O enquadramento pelo modo de produção, o tratamento do Design Participativo e a busca por trabalhos latino-americanos vieram de pedidos do autor ao longo do diálogo. O autor assume total responsabilidade pelo conteúdo.

Esta declaração faz parte do método e não é só uma formalidade. Um artigo que critica a substituição do trabalho de pesquisa por LLMs foi produzido com um LLM. A Seção~\ref{pt-reflexiva} discute o que isso significa.

\subsection{Verificação e limitações}

Cada referência foi conferida em pelo menos um registro primário: página de editora, repositório ou base bibliográfica. Uma segunda conferência, em novembro de 2026, reexaminou as 42 referências. Cada uma tem agora DOI, link de editora ou repositório, ou ISBN com registro em catálogo. Uma afirmação que não pôde ser rastreada até a fonte foi retirada. Os livros são citados por sua tese central, sem número de página. Uma referência é um preprint ainda sem revisão por pares (ROTHSCHILD \textit{et al.}, 2025), e o texto o indica.

A revisão tem limites que o leitor deve considerar. Muitas fontes foram avaliadas por resumos, sínteses das editoras e textos parciais, e não por leitura integral. As afirmações sobre essas fontes ficam dentro do que esse material sustenta. As buscas em espanhol retornaram sobretudo material não acadêmico. A revisão não cobriu tradições como a ergonomia da atividade, que poderiam oferecer outras críticas ao usuário simulado.

\section{O problema: produção, conhecimento e participação}\label{pt-problema}

\subsection{Uma necessidade do modo de produção}

Na produção de software, o teste de usabilidade é uma etapa do ciclo que leva o produto da concepção ao mercado. É uma etapa lenta, porque depende do tempo de outras pessoas.

Sob a concorrência capitalista, o tempo entre conceber e vender é custo. A empresa que lança antes chega antes ao mercado. Cada etapa que não pode ser acelerada vira um obstáculo à valorização do capital. A pesquisa com usuários é uma dessas etapas. Ela não acelera no ritmo do código, porque a pessoa consultada não trabalha no ritmo da máquina. \reading{Esta é uma leitura deste artigo.} A literatura técnica nomeia o mesmo fato com outras palavras. Ivory e Hearst (2001, p. 470) justificaram a avaliação automática de usabilidade porque a avaliação ``can be expensive in terms of time and human resources'', isto é, pode ser cara em tempo e em recursos humanos. Agnew \textit{et al.} (2024) identificaram a redução do custo da pesquisa como uma das principais motivações das propostas de substituir participantes humanos por IA generativa.

Os usuários sintéticos respondem diretamente a essa pressão. Eles trabalham no ritmo da máquina, a qualquer hora e em qualquer quantidade.

\subsection{A separação entre concepção e execução}

Braverman (1974) mostrou que o capitalismo monopolista reorganiza o processo de trabalho em dois movimentos. O primeiro separa a concepção do trabalho de sua execução. O segundo retira do trabalhador o saber sobre o ofício e o transfere para a gerência e para a máquina. Qin e Cheon (2026) propuseram trazer a teoria do processo de trabalho para a IHC. Eles examinam a literatura de IHC e de trabalho cooperativo apoiado por computador sobre trabalho, e pedem que o campo trate o trabalho como questão de economia política, e não apenas como execução de tarefas com tecnologia.

\reading{Este artigo lê o usuário simulado como repetição dos dois movimentos de Braverman, em dois planos.} No primeiro plano, ele retira o saber do participante. O que uma pessoa sabe sobre como usa, evita e contorna uma interface vira um padrão estatístico extraído de textos que ela escreveu com outros fins. No segundo plano, ele retira o saber do pesquisador de UX. Conduzir uma sessão, escutar, ler o silêncio e a hesitação viram um procedimento automático. O que antes exigia um ofício passa a exigir uma assinatura de serviço.

O movimento também se reproduz. A pesquisa com usuários já tinha passado por uma primeira mercantilização com as plataformas de trabalho por tarefa, como o Mechanical Turk, da Amazon. Gray e Suri (2019) descrevem o trabalho humano invisível por trás desses serviços automatizados. Kapania \textit{et al.} (2025) lembram que os LLMs foram treinados com textos de pessoas que não autorizaram esse uso e com o trabalho de anotadores de dados mal remunerados. A simulação dessas pessoas é então oferecida como produto. Westwood (2025) mostrou que um respondente sintético autônomo, construído com um LLM, produziu respostas coerentes em pesquisas online e escapou da maior parte dos métodos de detecção em uso. A técnica que promete substituir o participante ameaça, assim, os dados dos participantes reais. Parte dos autores do UXAgent, sistema discutido na Seção~\ref{pt-abordagens}, trabalha na Amazon, a empresa que opera o Mechanical Turk (LU \textit{et al.}, 2025).

\subsection{Os conflitos com a pesquisa acadêmica}

A pesquisa acadêmica e a produção de mercadorias perguntam coisas diferentes aos usuários. A empresa quer saber se o produto funciona o bastante para ser vendido. A pesquisa quer saber como as pessoas vivem com a tecnologia e o que essa experiência ensina. \reading{Na leitura deste artigo, o usuário simulado atende melhor à primeira pergunta do que à segunda.}

O conflito aparece em pelo menos quatro pontos.

O primeiro é o valor do conhecimento. Agnew \textit{et al.} (2024) argumentam que as propostas de substituição contrariam os valores que fundam a pesquisa com participantes humanos: representação, inclusão e compreensão. Esses valores não entram no cálculo de custo de um ciclo de produto.

O segundo é a legitimidade dos métodos. Kapania \textit{et al.} (2025) registram que pesquisadores qualitativos temem a deslegitimação dos seus métodos. Se uma entrevista sintética custa centavos e responde na hora, a entrevista com uma pessoa passa a parecer um luxo injustificável.

O terceiro é a própria universidade. \reading{Aqui o artigo volta a propor uma leitura.} A pressão por produtividade acadêmica, medida em número de artigos, empurra pesquisadores para os mesmos atalhos que a indústria busca. Parte relevante da literatura sobre usuários sintéticos vem de laboratórios ligados a grandes empresas de tecnologia, como Amazon e Google (LU \textit{et al.}, 2025; DUAN \textit{et al.}, 2024a). A fronteira entre pesquisa e desenvolvimento de produto fica menos nítida.

O quarto é a ética em pesquisa. No Brasil, a Resolução nº 510/2016 do Conselho Nacional de Saúde (CNS) protege participantes humanos nas Ciências Humanas e Sociais (BRASIL, 2016). Um estudo feito só com usuários sintéticos não tem participantes humanos no sentido da resolução. Ele escapa à proteção, embora produza decisões que afetam pessoas. A comunidade brasileira de IHC incluiu a ética da participação humana na pesquisa entre os seus grandes desafios para 2025--2035 (PEREIRA; DARIN; SILVEIRA, 2024). A Seção~\ref{pt-etica} retoma esse ponto.

\subsection{O Design Participativo como caso-limite}\label{pt-dp}

O Design Participativo torna o conflito mais nítido do que qualquer outro método. Ele nasceu na Escandinávia, nos anos 1970, em projetos com sindicatos que discutiam a entrada de computadores nas fábricas e nos escritórios. Ehn (1992) descreve essa origem como uma luta por democracia no local de trabalho. Leis da época, como a lei sueca de codeterminação, criaram condições para que os sindicatos influenciassem os sistemas que mudariam o trabalho dos seus membros.

Kensing e Blomberg (1998) organizam o campo em torno de três questões: a política do design, a natureza da participação e os métodos, ferramentas e técnicas que tornam a participação possível. A primeira questão define as outras duas. No Design Participativo, as pessoas não participam para fornecer dados ao designer. Elas participam porque têm o direito de influir em uma tecnologia que vai mudar a sua vida.

A América Latina tem uma raiz própria para essa ideia. A investigação-ação participativa, sistematizada por Fals Borda (1987), trata a pesquisa como produção conjunta de conhecimento com as comunidades, orientada para a transformação social. Escobar (2018) estende o argumento ao design para a autonomia, no qual as comunidades projetam as condições da própria vida. Lima e Lima (2025) propuseram recentemente um Design Participativo sensível ao contexto brasileiro, apoiado na terceira onda da IHC e no pensamento decolonial.

Essa premissa torna o participante simulado uma contradição em termos. Um agente pode responder perguntas. Ele não pode reivindicar, negociar, recusar ou se organizar. Ele não tem interesses a defender. \reading{No Design Participativo, simular o participante elimina o sujeito político do método e mantém apenas a sua aparência.}

Ehn (1992) acrescenta um argumento epistemológico. Apoiado em Wittgenstein, ele sustenta que a habilidade prática é tácita e não pode ser formalizada por inteiro. Por isso, o Design Participativo trabalha com protótipos, maquetes e experiência compartilhada, para que designers e usuários aprendam uns com os outros. O usuário simulado só conhece o que alguém já escreveu. O saber tácito, por definição, é o que não foi escrito.

\reading{O mesmo raciocínio vale, em graus diferentes, para outros métodos que dependem da presença da pessoa.} A etnografia depende do pesquisador estar no lugar onde a tecnologia é usada. A entrevista qualitativa depende da relação entre quem pergunta e quem responde (KAPANIA \textit{et al.}, 2025). O teste de acessibilidade depende das pessoas com deficiência, que estão nas bordas das distribuições que os modelos tendem a apagar (WANG; MORGENSTERN; DICKERSON, 2025). Quanto mais um método trata o participante como sujeito, e não como fonte de dados, menos espaço deixa para a simulação. O Design Participativo ocupa o extremo dessa escala. Nele, a IA generativa pode ajudar a preparar materiais ou a ensaiar a facilitação. Ela não pode ocupar a cadeira de quem participa.

\section{Uma genealogia do usuário simulado}\label{pt-genealogia}

\subsection{O usuário idealizado: os modelos cognitivos}

A primeira forma do usuário simulado surgiu no centro de pesquisa da Xerox em Palo Alto (PARC). Em 1983, Card, Moran e Newell publicaram \textit{The Psychology of Human-Computer Interaction}, que propunha uma psicologia aplicada aos usuários de sistemas interativos (CARD; MORAN; NEWELL, 1983). O livro trouxe dois instrumentos. O modelo GOMS descreve a tarefa em metas, operadores, métodos e regras de seleção. O modelo de nível de teclas (KLM, do inglês \textit{keystroke-level model}) atribui um tempo a cada operação elementar e soma esses tempos para prever a duração da tarefa.

O usuário desses modelos é idealizado. O KLM prevê o tempo de um especialista que executa uma tarefa rotineira sem erros. Aprendizagem, erro e cansaço ficam de fora. Uma teoria explícita da mente compensava a ausência da pessoa.

Duas décadas depois, John \textit{et al.} (2004) observaram que esses modelos tinham pouco impacto na prática, porque construí-los era trabalhoso. Os autores criaram o CogTool. Nele, o designer demonstra a tarefa em um protótipo HTML e o sistema gera o modelo automaticamente. O modelo roda na arquitetura cognitiva ACT-R, que simula percepção e movimento, e interage com o protótipo. O CogTool é o antecedente técnico mais direto dos agentes atuais: um usuário simulado que opera uma interface e produz dados de desempenho.

\subsection{O usuário interpretado: a inspeção por especialistas}

A segunda forma vem da engenharia de usabilidade de baixo custo. Nielsen e Molich (1990) apresentaram a avaliação heurística, na qual avaliadores examinam a interface com base em uma lista curta de princípios. Eles mostraram que um avaliador isolado encontrava entre 20\% e 51\% dos problemas, e que o trabalho somado de três a cinco avaliadores alcançava bons resultados. No mesmo artigo, consideraram a avaliação automática inviável, salvo em verificações pontuais.

A avaliação heurística substitui o usuário por um intérprete do usuário. A IA generativa, como mostra a Seção~\ref{pt-abordagens}, automatiza esse intérprete.

Ivory e Hearst (2001) organizaram os métodos de avaliação em uma taxonomia centrada na automação. As autoras viam a automação como forma de ampliar os métodos existentes, justificada pelo custo do teste com pessoas. O argumento do custo atravessa toda esta genealogia.

\subsection{O conflito fundador: a ação situada}

O mesmo laboratório que produziu o GOMS produziu sua crítica mais forte. Em \textit{Plans and Situated Actions}, Suchman (1987) estudou pessoas usando uma copiadora com um sistema de ajuda desenhado segundo o modelo de planejamento da ciência cognitiva. A interação raramente seguia o plano, nem o da pessoa nem o da máquina. Para Suchman, os planos são recursos vagos para lidar com as contingências da situação, e não programas que determinam a ação. O problema da ciência cognitiva estava em trocar a ação pelo plano e a circunstância real pela representação da circunstância.

Essa crítica atinge todas as formas do usuário simulado. Um modelo GOMS é um plano formalizado. Um agente generativo também opera por planejamento, memória e reflexão. Em ambos os casos, a situação de uso entra no teste apenas como representação.

\subsection{O usuário estatístico: a virada dos dados}

Entre os modelos cognitivos e os LLMs há uma etapa intermediária. O modelo do usuário deixa de vir de uma teoria e passa a vir de dados de comportamento humano. Swearngin e Li (2019) reuniram, por colaboração coletiva, rótulos sobre quais elementos de interfaces móveis as pessoas percebiam como tocáveis. Com esses rótulos, treinaram uma rede neural que prevê essa percepção, com precisão média de 90,2\% e revocação de 87,0\%. A ferramenta aponta divergências entre a intenção do designer e o que os usuários provavelmente vão perceber.

O usuário dessa etapa é uma média de milhares de pessoas reais. A tarefa simulada é estreita e mensurável, e a validação compara o modelo com pessoas. No Brasil, Iannuzzi \textit{et al.} (2025) mapearam essa linha de trabalho. Dos 238 estudos encontrados, analisaram 47. Os métodos mais comuns aplicam modelos supervisionados a métricas de interação, verbalizações de protocolos de pensar em voz alta e imagens de interface. Os autores apontam limites na generalização dos modelos e na integração com especialistas em IHC e UX.

\subsection{O usuário crível: os simulacros generativos}

A forma atual do usuário simulado ganhou nome em dois trabalhos do grupo de Joon Sung Park, em Stanford. Em \textit{Social Simulacra}, Park \textit{et al.} (2022) geraram milhares de membros de uma comunidade online e suas interações a partir de uma descrição do designer, com metas, regras e personas. A técnica se apoiava em uma observação explícita: os dados de treino do modelo já contêm grande variedade de comportamentos de redes sociais. Nas avaliações, os participantes muitas vezes não distinguiram as interações simuladas das reais. Em \textit{Generative Agents}, Park \textit{et al.} (2023) criaram agentes com memória, reflexão e planejamento, que viviam em uma cidade virtual de 25 habitantes. O critério de avaliação declarado foi a credibilidade do comportamento.

A palavra simulacro define o conceito. O agente não pretende prever o que uma pessoa faria. Ele pretende produzir um comportamento que pareça humano.

\subsection{Uma retomada: a racionalidade computacional}

Nem todo trabalho recente segue o caminho generativo. Oulasvirta, Jokinen e Howes (2022) retomam explicitamente a pergunta de Card, Moran e Newell sobre como as pessoas interagem com computadores. Eles a reformulam como questão de adaptação: os usuários agem do melhor modo possível dentro dos limites da sua cognição, do dispositivo e do ambiente. Essa teoria, chamada racionalidade computacional, pode ser expressa em modelos que explicam e preveem a interação.

\reading{Na leitura deste artigo, essa retomada funciona como contraponto silencioso à simulação com LLMs.} Ela mantém o projeto de simular o usuário, mas exige uma teoria explícita do comportamento. O agente generativo simula sem essa teoria. Seu comportamento vem da estatística do texto, e não de uma hipótese sobre a mente. A genealogia se divide aqui entre um ramo que modela a cognição e outro que modela a linguagem sobre a cognição.

\subsection{O deslocamento do critério de validade}\label{pt-deslocamento}

\reading{A genealogia revela uma mudança no que conta como um bom modelo de usuário. Esta é a principal tese analítica do artigo.} Os modelos de Card, Moran e Newell (1983) e de John \textit{et al.} (2004) eram julgados pela precisão: o tempo previsto precisava coincidir com o medido. O modelo de Swearngin e Li (2019) também era julgado pela concordância com pessoas. Os simulacros de Park \textit{et al.} (2022; 2023) trocaram a precisão pela credibilidade.

Essa troca tem consequências. Um usuário idealizado e preciso diz algo verificável sobre um tipo restrito de tarefa. Um usuário crível e impreciso diz algo plausível sobre qualquer tarefa. O segundo cobre um território muito maior e é muito mais difícil de refutar. A Tabela~\ref{pt-tab-genealogia} resume a sequência.

\begin{table}[ht]
\centering\small
\caption{Formas do usuário simulado na IHC.}\label{pt-tab-genealogia}
\begin{tabularx}{\textwidth}{@{}p{2.1cm}LLLL@{}}
\toprule
\textbf{Período} & \textbf{Figura do usuário} & \textbf{Fonte do modelo} & \textbf{Critério de validade} & \textbf{Obras centrais} \\
\midrule
Anos 1980--2000 & Especialista idealizado, sem erros & Teoria da mente (GOMS, KLM, ACT-R) & Precisão do tempo previsto & Card, Moran e Newell (1983); John \textit{et al.} (2004) \\
Anos 1990--2000 & Interpretado por um especialista & Princípios de design aplicados por avaliadores & Número de problemas encontrados & Nielsen e Molich (1990); Ivory e Hearst (2001) \\
Anos 2010 & Média estatística de pessoas reais & Dados de comportamento e aprendizado de máquina & Concordância com rótulos humanos & Swearngin e Li (2019); Iannuzzi \textit{et al.} (2025) \\
Desde 2022 & Simulacro crível & Estatística do texto em LLMs & Credibilidade para observadores & Park \textit{et al.} (2022; 2023); Lu \textit{et al.} (2025) \\
Desde 2022 & Agente adaptativo & Teoria da adaptação limitada & Explicação e previsão & Oulasvirta, Jokinen e Howes (2022) \\
\bottomrule
\end{tabularx}
\end{table}

\section{Abordagens atuais e seus conflitos}\label{pt-abordagens}

\subsection{O modelo como avaliador}

Na primeira abordagem, o LLM ocupa o lugar do especialista da avaliação heurística. Duan \textit{et al.} (2024a) testaram o GPT-4 em um plugin para o Figma, que recebe a tela e as diretrizes e devolve sugestões. Avaliaram 51 interfaces com três conjuntos de diretrizes, compararam os resultados com os de especialistas humanos e estudaram a ferramenta com 12 designers experientes. O modelo detectou erros sutis e melhorou textos da interface, mas suas sugestões perderam utilidade ao longo das iterações. O plugin lia a tela como uma estrutura de dados em JSON, uma tela estática de cada vez. Em trabalho posterior, o mesmo grupo reconheceu que as técnicas baseadas em LLM ainda não alcançam avaliadores humanos. O desempenho melhorou 55\% quando o modelo recebeu críticas de interface escritas por pessoas (DUAN \textit{et al.}, 2024b).

Estudos brasileiros apontam na mesma direção. Lima \textit{et al.} (2025) compararam quatro sistemas de IA generativa na avaliação dos sites da COP30 e da COP29. Os sistemas encontraram problemas de navegação, de acessibilidade e de terminologia, mas mostraram limites em sensibilidade cultural e em acessibilidade técnica. Em versão ampliada, Lima \textit{et al.} (2026) compararam as avaliações dos sistemas com as percepções de dez participantes humanos. As pessoas relataram alta satisfação com o site da COP30, enquanto os sistemas de IA o penalizaram por deficiências estruturais. Os dois grupos concordaram sobre o critério de pior desempenho. Os autores defendem modelos híbridos, nos quais a IA fornece dados heurísticos sistemáticos e as pessoas acrescentam interpretação prática e sociocultural.

Matthaus \textit{et al.} (2026) estudaram a mesma abordagem na educação. Em um experimento com 24 estudantes de engenharia de software, não encontraram diferença estatisticamente significativa entre os grupos com e sem IA generativa, mas o grupo com IA teve média menor de identificações corretas. Os registros de interação mostraram uso sobretudo passivo, aceitação acrítica das respostas, alucinações, respostas bajuladoras e dificuldade de perceber os erros da ferramenta. Os autores alertam para a delegação cognitiva na aprendizagem de conceitos.

\subsection{O modelo como participante}

Na segunda abordagem, o LLM ocupa o lugar do participante. Hämäläinen, Tavast e Kunnari (2023) usaram o GPT-3 para gerar respostas abertas de questionário sobre a experiência de videogames como arte. Leitores tiveram dificuldade em distinguir respostas sintéticas de reais. Os autores recomendaram o uso desses dados em ideação e pilotos, sempre com validação posterior em dados reais.

Nas ciências sociais, Argyle \textit{et al.} (2023) propuseram o conceito de fidelidade algorítmica. Segundo essa visão, o viés do modelo é fino e ligado a grupos demográficos, de modo que condicionar o modelo com histórias de vida reproduz as distribuições de resposta de subgrupos humanos. É a defesa mais forte da substituição.

O sistema UXAgent leva essa lógica ao teste de usabilidade (LU \textit{et al.}, 2025). Ele gera milhares de personas e coloca cada uma para navegar no site avaliado, com registro de ações, vídeo e entrevista com o agente. Os autores definem o uso como ensaio do protocolo de pesquisa antes do estudo com pessoas. Os cinco pesquisadores de UX que avaliaram o sistema elogiaram a proposta e também expressaram preocupação com o futuro da pesquisa de UX feita com agentes.

\subsection{Conflitos epistemológicos}\label{pt-epistemicos}

Conflito epistemológico, neste artigo, é uma tensão sobre a validade do conhecimento produzido. A pergunta é se o que a simulação produz pode valer como evidência sobre pessoas.

\paragraph{Credibilidade não é precisão.} O grupo do UXAgent mediu se os agentes reproduzem ações reais em 31.865 sessões de compra online, com 230.965 ações de usuários. Os modelos acertaram 11,86\% das ações humanas (LU \textit{et al.}, 2026). O ajuste fino com dados reais de cliques melhorou o resultado. O estudo recoloca a precisão como critério e, nos termos desta genealogia, retoma o padrão dos modelos cognitivos. A comparação brasileira de Lima \textit{et al.} (2026) mostra a mesma distância na abordagem do avaliador: o julgamento da IA e a experiência dos usuários divergiram.

\paragraph{A média apaga a variação.} Bisbee \textit{et al.} (2024) mostraram que as médias sintéticas se aproximam das reais em pesquisas de opinião, mas as respostas sintéticas variam menos que as humanas e os coeficientes de regressão divergem. O estudo contesta a fidelidade algorítmica. Wang, Morgenstern e Dickerson (2025) testaram quatro LLMs contra 3.200 participantes de 16 identidades demográficas. Concluíram que os LLMs retratam mal e achatam os grupos que simulam, e que os comandos de identidade tendem a essencializar as identidades. Eles explicam o achatamento por duas causas: o treinamento recompensa a saída mais provável, e o texto coletado raramente vem ligado à identidade de quem o escreveu. Na usabilidade, os problemas aparecem com frequência nas bordas da distribuição, como a pessoa idosa, a pessoa com baixa visão ou a pessoa com pouca experiência digital. \reading{Um método que comprime a variação tende a esconder esses casos.}

\paragraph{Falta a situação.} Kapania \textit{et al.} (2025) entrevistaram 19 pesquisadores qualitativos. As respostas sintéticas careciam de palpabilidade e de profundidade de contexto. Os autores chamam o fenômeno de efeito substituto. \reading{Essa crítica repete o argumento de Suchman (1987) quase quatro décadas depois.} A usabilidade se avalia no encontro entre um corpo situado e um artefato. O agente não tem corpo, cansaço, pressa nem história com a tecnologia. Ele lê a interface como texto estruturado.

\paragraph{Os resultados não se repetem.} Bisbee \textit{et al.} (2024) mostraram que pequenas mudanças na redação do comando alteram as respostas, e que o mesmo comando produziu resultados diferentes ao longo de três meses. As empresas atualizam os modelos sem aviso.

\paragraph{O dado real se contamina.} Westwood (2025) mostrou que a coerência deixou de indicar autoria humana em pesquisas online. Os testes remotos de usabilidade não moderados, comuns na prática de UX, entram na mesma zona de risco. Em preprint, Rothschild \textit{et al.} (2025) responderam que as evidências ainda não mostram que a contaminação seja real e grande, e o tamanho da ameaça segue em debate.

\paragraph{O teste confirma o que já se sabe.} \reading{Esta é uma hipótese do artigo.} O modelo aprende com textos que descrevem como as pessoas usam interfaces, incluindo diretrizes e boas práticas. Ao simular um usuário, tende a devolver esse saber já escrito. A queda de utilidade ao longo das iterações relatada por Duan \textit{et al.} (2024a) é compatível com essa hipótese, embora os autores não a formulem assim.

\subsection{Conflitos éticos}\label{pt-etica}

\paragraph{Consentimento e agência.} Kapania \textit{et al.} (2025) registram que os LLMs suprimem o consentimento e a agência dos participantes. O participante humano pode recusar, interromper e contestar a pergunta. O agente não faz nada disso por vontade própria. A Resolução nº 510/2016 define o consentimento livre e esclarecido como a anuência do participante ``livre de simulação, fraude, erro ou intimidação'' (BRASIL, 2016, art. 2º, V).

\paragraph{Uma lacuna regulatória.} A resolução se aplica a pesquisas que usam dados obtidos diretamente com participantes, informações identificáveis ou procedimentos com risco maior que o da vida cotidiana (BRASIL, 2016, art. 1º). Um estudo feito só com usuários sintéticos não se encaixa em nenhuma dessas condições. Ele tende a ficar fora do sistema formado pelos Comitês de Ética em Pesquisa (CEP) e pela Comissão Nacional de Ética em Pesquisa (CONEP). \reading{Na leitura deste artigo, o resultado é um estudo que fala sobre pessoas e orienta decisões de produto que as afetam, sem passar por nenhuma instância de proteção.} A literatura internacional revisada não discute essa lacuna, porque ela depende do marco normativo brasileiro.

\paragraph{A ilusão de inclusão.} Agnew \textit{et al.} (2024) concluem que as propostas de substituição prometem diversidade sem a presença de pessoas diversas. A participação também é o meio pelo qual as pessoas influenciam as tecnologias que vão usar.

\paragraph{Dano a grupos marginalizados.} Wang, Morgenstern e Dickerson (2025) ligam o achatamento a um histórico de retratos unidimensionais de grupos marginalizados. Pedem cautela sempre que a identidade do participante for relevante para a tarefa. Em testes de acessibilidade, ela quase sempre é.

\paragraph{Dados de treino e trabalho invisível.} Como visto na Seção~\ref{pt-problema}, o usuário sintético carrega a voz não consentida de pessoas reais e o trabalho de anotadores de dados (KAPANIA \textit{et al.}, 2025; GRAY; SURI, 2019).

\paragraph{O argumento da proteção.} O conflito ético tem dois lados. Lu \textit{et al.} (2025) defendem que simular antes de testar protege os participantes de protocolos mal desenhados. O argumento vale para o piloto. Ele se enfraquece quando o piloto passa a ocupar o lugar do estudo, que é o que a motivação de custo tende a produzir.

A Tabela~\ref{pt-tab-conflitos} resume os conflitos por tipo de uso.

\begin{table}[ht]
\centering\small
\caption{Conflitos por tipo de uso da simulação com LLMs.}\label{pt-tab-conflitos}
\begin{tabularx}{\textwidth}{@{}p{2.8cm}LLp{3.2cm}@{}}
\toprule
\textbf{Uso} & \textbf{Conflito epistemológico} & \textbf{Conflito ético} & \textbf{Fontes} \\
\midrule
Avaliação heurística automatizada & Desempenho inferior ao humano; tela lida como dados; possível circularidade & Baixo com revisão humana; risco de delegação cognitiva no ensino & Duan \textit{et al.} (2024a; 2024b); Lima \textit{et al.} (2025; 2026); Matthaus \textit{et al.} (2026) \\
Piloto de protocolo & Baixo, se o achado não valer como evidência sobre usuários & Pode proteger participantes reais & Lu \textit{et al.} (2025); Hämäläinen, Tavast e Kunnari (2023) \\
Teste com usuários sintéticos & Credibilidade sem precisão; variação achatada; baixa reprodutibilidade & Sem consentimento; fora do sistema CEP/CONEP & Lu \textit{et al.} (2026); Bisbee \textit{et al.} (2024); Brasil (2016) \\
Personas de grupos específicos & Retrato feito de fora do grupo; identidade essencializada & Dano representacional & Wang, Morgenstern e Dickerson (2025) \\
Entrevistas sintéticas & Falta de palpabilidade e de contexto & Agência suprimida; dados de treino não consentidos & Kapania \textit{et al.} (2025) \\
Substituição ampla & Fidelidade algorítmica contestada & Ilusão de inclusão & Argyle \textit{et al.} (2023); Agnew \textit{et al.} (2024) \\
Coleta online não moderada & Contaminação do dado humano & Confiança na base de evidências & Westwood (2025) \\
Design Participativo & Saber tácito fora de alcance & Sujeito político eliminado & Ehn (1992); Kensing e Blomberg (1998) \\
\bottomrule
\end{tabularx}
\end{table}

\section{O olhar da América Latina}\label{pt-latam}

\subsection{O que a região produz}

A produção latino-americana encontrada nesta revisão se concentra na abordagem do avaliador. Estudos brasileiros comparam a IA generativa com avaliadores e usuários humanos (LIMA \textit{et al.}, 2025; 2026), mapeiam o aprendizado de máquina na avaliação de usabilidade (IANNUZZI \textit{et al.}, 2025) e estudam os efeitos da IA generativa na aprendizagem da avaliação heurística (MATTHAUS \textit{et al.}, 2026). No Peru, Lecaros \textit{et al.} (2024) automatizaram o método heurístico de Granollers com um sistema de apoio que não usa IA. O padrão comum trata a IA como apoio ao julgamento humano e defende arranjos híbridos.

\subsection{Uma ausência}

A revisão não encontrou nenhum estudo acadêmico latino-americano, em português ou em espanhol, sobre usuários sintéticos no lugar de participantes de testes de UX. As buscas em espanhol retornaram textos de blogs e material de marketing de empresas que vendem o serviço. \reading{Na leitura deste artigo, a ausência é ela mesma um achado.} As ferramentas chegam à região como produtos antes de chegar como objetos de pesquisa. Se os modelos retratam grupos de fora (WANG; MORGENSTERN; DICKERSON, 2025), o risco cresce para populações sub-representadas nos dados de treino. Ricaurte (2019) trata as epistemologias centradas em dados como expressão da colonialidade do poder. Faustino e Lippold (2023) analisam o colonialismo digital numa perspectiva fanoniana. Os dois referenciais sugerem que um usuário latino-americano produzido por um modelo treinado sobretudo com textos do Norte global não é uma aproximação neutra.

\subsection{Recursos da região}

A região também oferece recursos para enfrentar o problema. A comunidade brasileira de IHC definiu sete grandes desafios para 2025--2035. Três deles tocam diretamente este artigo: ética e responsabilidade na participação humana em pesquisas de IHC (GC2), pluralidade e decolonialidade na IHC (GC3) e implicações da inteligência artificial na IHC (GC6) (PEREIRA; DARIN; SILVEIRA, 2024). Martins, Lana e Villela (2025) responderam ao desafio da IA estudando como os valores éticos são comunicados nas interfaces de sistemas de IA generativa.

A engenharia semiótica, desenvolvida no Brasil, trata a interface como uma mensagem dos designers para os usuários (DE SOUZA, 2005). Seu método de avaliação de comunicabilidade depende de observar a interação dos usuários e identificar momentos de ruptura na comunicação (PRATES; DE SOUZA; BARBOSA, 2000). \reading{Na leitura deste artigo, um usuário sintético pode produzir uma ruptura prevista, mas não uma ruptura vivida por alguém que não entendeu a mensagem do designer.} O método oferece, portanto, uma razão de princípio para manter as pessoas na avaliação.

Álvaro Vieira Pinto (2005) criticou a ideia de uma ``era tecnológica'' como ideologia e tratou a tecnologia como produção humana situada na história. \reading{Lido a partir de sua obra, o usuário sintético aparece menos como etapa inevitável do progresso técnico e mais como escolha feita sob condições específicas de produção.} A investigação-ação participativa (FALS BORDA, 1987) e o design para a autonomia (ESCOBAR, 2018), discutidos na Seção~\ref{pt-dp}, completam esse conjunto de recursos regionais.

\section{Discussão}\label{pt-discussao}

\subsection{A credibilidade como produto}

\reading{Esta seção desenvolve a interpretação do artigo.} O deslocamento da precisão para a credibilidade (Seção~\ref{pt-deslocamento}) e a pressão da produção (Seção~\ref{pt-problema}) convergem. Uma resposta crível chega rápido e custa pouco. Uma resposta precisa exige pessoas, tempo e medição. Quando o critério de validade passa a ser a credibilidade, o modelo de usuário vira algo que pode ser vendido como serviço, porque a credibilidade é uma propriedade que o próprio modelo consegue gerar. A precisão, ao contrário, só pode ser estabelecida diante de pessoas. Os estudos que voltam à precisão (LU \textit{et al.}, 2026; BISBEE \textit{et al.}, 2024) e a comparação brasileira entre o julgamento da IA e a experiência dos usuários (LIMA \textit{et al.}, 2026) mostram o custo dessa troca.

\subsection{Duas abordagens, dois pesos}

Automatizar o avaliador herda os limites da avaliação heurística, que sempre foi um método sem usuário. Automatizar o participante é uma ruptura maior, porque elimina a fonte que o teste de usabilidade existe para consultar. A abordagem do avaliador também tem problemas. Matthaus \textit{et al.} (2026) mostram que, na educação, ela pode trocar a aprendizagem da avaliação pela delegação da avaliação. \reading{Nos termos de Braverman, a primeira abordagem arrisca desqualificar o avaliador, enquanto a segunda elimina o participante.}

\subsection{Uma nota reflexiva sobre este artigo}\label{pt-reflexiva}

Como informa a Seção~\ref{pt-uso-ia}, este artigo foi produzido com um LLM como assistente de pesquisa. \reading{Esse processo é, ele mesmo, um caso da separação entre concepção e execução discutida na Seção~\ref{pt-problema}.} O modelo executou tarefas que um assistente de pesquisa humano executaria: buscar, resumir, conferir registros, redigir e traduzir. Ele não definiu a pergunta de pesquisa nem o enquadramento teórico. O enquadramento pelo modo de produção, o tratamento do Design Participativo e a busca por fontes latino-americanas vieram de pedidos do autor em diferentes momentos do diálogo.

O processo também produziu erros que a verificação corrigiu. Um rascunho inicial afirmava que a Amazon tinha financiado o UXAgent. As fontes sustentam apenas que parte dos autores trabalha na Amazon, e o texto foi corrigido. As datas de acesso de uma versão inicial das referências estavam inconsistentes e também foram corrigidas. Esses casos sustentam o argumento do artigo de modo modesto. O texto produzido por um LLM é crível antes de ser verificado. Sua precisão precisou ser estabelecida pela conferência de cada afirmação em registros primários, o mesmo movimento que Lu \textit{et al.} (2026) fazem com os usuários simulados.

O artigo não afirma que esse modo de produção é neutro. Ele o relata para que os leitores possam pesar o trabalho de acordo.

\section{Conclusões}\label{pt-conclusoes}

A primeira conclusão é que o usuário simulado é uma figura antiga da IHC. Ele mudou de forma conforme a técnica disponível. Foi um especialista idealizado nos anos 1980, um intérprete do usuário na avaliação heurística, uma média estatística na década de 2010 e um simulacro crível a partir de 2022. Cada forma herdou o argumento do custo e a objeção de Suchman sobre a situação.

A segunda conclusão é que a mudança decisiva não foi técnica. Foi a troca da precisão pela credibilidade como critério de validade. Essa troca permite que o usuário simulado fale sobre qualquer tarefa, ao custo de dizer pouco que se possa verificar. Os estudos recentes que voltam à precisão encontram resultados fracos.

A terceira conclusão é que a literatura converge em um consenso mínimo. A simulação pode ajudar a ensaiar protocolos e a fazer uma triagem inicial de problemas. Ela não substitui o teste com pessoas. A divergência real está na natureza dos limites. Os trabalhos técnicos tratam as falhas como lacunas que mais dados e ajuste fino vão fechar. Os trabalhos críticos tratam as falhas como propriedades do método. Um modelo treinado para produzir a saída mais provável não pode representar a pessoa improvável, que é muitas vezes quem revela o problema da interface.

A quarta conclusão é que as duas abordagens têm pesos diferentes. Automatizar o avaliador arrisca a desqualificação. Automatizar o participante elimina a fonte de conhecimento que o teste existe para alcançar. No Design Participativo, elimina o sujeito político do método.

A quinta conclusão diz respeito à América Latina. A região usa a IA generativa como apoio à inspeção e ainda não estudou os usuários sintéticos. Essa lacuna importa, porque as pessoas mais sujeitas a ser mal retratadas por modelos treinados em outros lugares são também as menos presentes na pesquisa sobre esses modelos.

\section{Problemas em aberto e trabalhos futuros}\label{pt-futuro}

\paragraph{Ética e regulação no Brasil.} Falta examinar como o sistema CEP/CONEP deve tratar estudos que usam apenas usuários sintéticos para decidir sobre produtos usados por pessoas. Uma linha de trabalho é documental: analisar pareceres, resoluções e consultas sobre o tema. Outra é propositiva: formular critérios de declaração e de revisão para esse tipo de estudo.

\paragraph{Populações do Sul global.} Não foi encontrado nenhum estudo empírico sobre usuários simulados com populações brasileiras ou latino-americanas. Uma replicação de Wang, Morgenstern e Dickerson (2025) com grupos brasileiros, variando região, renda e letramento digital, testaria se o achatamento é maior fora dos centros onde os dados de treino são produzidos.

\paragraph{Corpo e situação.} A literatura aponta a falta de contexto e de palpabilidade, mas não desenvolve esse ponto como problema fenomenológico. Um trabalho teórico pode articular Suchman (1987) com a fenomenologia do corpo para explicar por que o encontro entre corpo e artefato não se reduz a texto.

\paragraph{Outras tradições críticas.} A ergonomia da atividade separa a tarefa prescrita da atividade real, e a engenharia semiótica trata a interface como comunicação. As duas podem oferecer contrapontos ao usuário simulado que esta revisão apenas começou a explorar.

\paragraph{Validação mista.} Faltam protocolos que combinem simulação e teste com pessoas e meçam o que cada um encontra. Um estudo comparativo pode registrar quais problemas só os agentes detectam, quais só as pessoas detectam e quais os dois detectam. Os testes de acessibilidade são um caso prioritário, porque concentram os usuários das bordas da distribuição.

\paragraph{Contaminação dos testes remotos.} Westwood (2025) estudou pesquisas de opinião. Falta medir a presença de agentes em testes de usabilidade remotos não moderados e avaliar métodos de verificação de autoria humana.

\paragraph{Circularidade.} A hipótese de que o usuário simulado devolve o saber já escrito sobre usabilidade pode ser testada. Um caminho é comparar o desempenho dos agentes em interfaces convencionais e em interfaces que rompem padrões conhecidos.

\paragraph{Reprodutibilidade.} Faltam padrões para registrar a versão do modelo, o comando, os parâmetros de amostragem e a data em estudos com usuários sintéticos, de modo que outros pesquisadores possam repeti-los.

\section*{Referências}
\addcontentsline{toc}{section}{Referências}

\end{document}